\documentclass[11pt]{article}
\usepackage[utf8]{inputenc} 
\usepackage{hyperref} 
\usepackage[a4paper, total={6.3in, 9.3in}]{geometry}
\usepackage{setspace}
\usepackage{graphicx}
\usepackage{subcaption}
\usepackage[english]{babel}
\usepackage{amsmath}
\usepackage{amssymb,bm}
\usepackage{cite}

\newcommand{\vv} {{\bm v}}

\newcommand{\lp}{\left(}
\newcommand{\rp}{\right)}
\newcommand{\lb}{\left[}
\newcommand{\rb}{\right]}
\newcommand{\lt}{\left}
\newcommand{\rt}{\right}
\newcommand{\la}{\left\langle}
\newcommand{\ra}{\right\rangle}
\newcommand{\pr}{\partial}
\newcommand{\al}{\alpha}
\newcommand{\bt}{\beta}
\newcommand{\prm}{\prime}
\newcommand{\bsl}{\boldsymbol}
\newcommand{\hf}{\frac{1}{2}}
\newcommand{\nn}{\nonumber} 
\newcommand{\om}{\omega}

\newcommand{\dd}{\textrm{d}}
\newcommand{\me}{\mathrm{e}}
\newcommand{\mc}{\mathcal}
\newcommand{\gm}{\gamma}
\newcommand{\gmm}{\Gamma}

\newcommand{\kp}{k^{\prime}}
\newcommand{\er}{Erd\H{o}s-R\'{e}nyi }

\newcommand{\bx}{{\bsl \xi}}
\newcommand{\prp}{\propto}
\newcommand{\be}{\begin{equation}}
\newcommand{\ee}{\end{equation}}
\newcommand{\bea}{\begin{eqnarray}}
\newcommand{\eea}{\end{eqnarray}}
\newcommand{\beas}{\begin{eqnarray*}}
\newcommand{\eeas}{\end{eqnarray*}}

\newcommand{\bbbone}{{\mathbb{I}}}

\begin{document}
\title{\bf A Random Walk Perspective on Hide-and-Seek Games}
\author{Shubham Pandey and Reimer K\"uhn\\
Mathematics Department, King's College London, Strand, London WC2R 2LS,UK\\[-2mm]
\date{\today}}
\maketitle

\begin{abstract}
We investigate hide-and-seek games on complex networks using a random walk framework. 
Specifically, we investigate the efficiency of various degree-biased random walk 
search strategies to locate items that are randomly hidden on a subset of vertices 
of a random graph. Vertices at which items are hidden in the network are chosen at 
random as well, though with probabilities that may depend on degree. We pitch various 
hide and seek strategies against each other, and determine the efficiency of search 
strategies by computing the average number of hidden items that a searcher will uncover 
in a random walk of $n$ steps. Our analysis is based on the cavity method for finite 
single instances of the problem, and generalises previous work of De Bacco et al.
\cite{debacco2015average} so as to cover degree-biased random walks. We also extend
the analysis to deal with the thermodynamic limit of infinite system size. We study a 
broad spectrum of functional forms for the degree bias of both the hiding and the search 
strategy and investigate the efficiency of families of search strategies for cases 
where their functional form is either matched or unmatched to that of the hiding 
strategy. Our results are in excellent agreement with those of numerical simulations.
We propose two simple approximations for predicting efficient search strategies. One 
is based on an equilibrium analysis of the random walk search strategy. While not 
exact, it produces correct orders of magnitude for parameters characterising optimal 
search strategies. The second exploits the existence of an effective drift in random
walks on networks, and is expected to be efficient in systems with low concentration of
small degree nodes.
\end{abstract}

\section{Introduction}

There are many real-life scenarios where one party attempts to hide information 
that is desired by another party. Examples include hiding confidential information 
at a node in a computer cluster, hiding an item at a physical site, and trying to 
keep an object safe from a potential attack on a group of sites. The party hiding 
the information could have good or bad intention.  E.g., they could be storing 
personal information, and the searcher could be a hacker trying to locate and exploit 
it; conversely, the hider could be stashing away stolen goods, and the seeker a 
police force looking for it. 

The above scenarios have been formalised by an agent-based game called hide-and-seek. 
In an abstract formulation, there are two agents, a hider and a searcher, and 
a sample space formalised as a network. The hider conceals a given set of items on 
a subset of vertices of the network. The searcher then tries to locate those objects 
by searching the network. 

Hide-and-seek games are a form of search games \cite{alpern2006theory, alpern2013search} 
-- a broad term that is used to describe games that involve an agent searching for 
something in a sample space. A multitude of search games have been explored over the years 
and they find applications in many fields \cite{alpern2006theory}. Search games can be 
applied for monitoring patrolling situations \cite{alpern2011patrolling}, controlling 
urban security \cite{tsai2010urban}, controlling contagion \cite{tsai2012security}, 
and detecting malicious packets in computer networks \cite{vanek}. Chapman et al. 
\cite{chapman2014hideandseek} study hide-and-seek games to address issues in cyber 
security, and we draw inspiration from their work. 

In the present paper we analyse hide-and-seek games from a random walk perspective.
I.e. we shall not be specifically concerned with game theoretic aspects, such as
existence or multiplicity of Nash equilibria. Rather we propose to analyse the 
{\em efficiencies\/} of families of random search strategies, formalised as degree-biased
random walks, when applied to locate a set of items hidden in a network according
to a probabilistic hiding strategy. We take the random hiding strategies to be degree-biased 
as well.

The problem of a random walker exploring a network has found applications in many 
fields \cite{tong2006fast}, including diffusion \cite{BrayRodg88}, infection dynamics 
in social networks \cite{More+02, New02}, or in image segmentation \cite{grady2006random}.

Random walks have been studied extensively over the years. Indeed, as emphasised 
by Lov\'asz \cite{lovasz1993random}, there is ``not much difference between the
theory of random walks on graphs and the theory of finite Markov chains", and 
so it will not come as a surprise that properties of random walks in complex 
networks have been studied in their own right \cite{lovasz1993random, NohRie04,
cooper2005cover, avin2008explore, metzler2014first, levin2017markov}.

In order to analyse the efficiency of a random search strategy, we adapt a result 
of De Bacco et al. \cite{debacco2015average} in which the average number of different 
vertices of a complex network visited by a random walker performing an unbiased $n$-step 
random walk is computed. We generalise their work by considering more general 
degree-biased transition probabilities and use this to assess the efficiency of a 
family of degree-biased random search algorithms to locate items that are randomly 
hidden on a subset of vertices of a random graph. We also extend the finite single
instance analysis of \cite{debacco2015average} to cover the thermodynamic limit
of infinite system size, using the method of \cite{kuehn2016} to isolate contributions
of the giant component of the system.

We compute, analyse and compare search efficiencies across a broad spectrum of 
search and hiding strategies. We explore a few basic types of graphs in the 
configuration model class, namely \er graphs, random regular graphs, and scale-free 
graphs. The principal reason for including the latter in our analysis is that 
many technical, social and biological real-life networks are indeed thought to 
be scale-free \cite{watts1998}. We compare our results with those obtained 
using random walk simulations on large finite instances and find them in excellent
agreement with those obtained using the theoretical tools developed in the present
paper.  

The remainder of our paper is organised as follows. In Sect. \ref{RWM}, we present 
the random walk framework in terms of which we are going to analyse the efficiency
of random search strategies, as well as the families of degree-biased strategies for 
hiding and searching we will investigate. In Sect. \ref{sec_theoryB} we describe the 
method that we use for computing the efficiency of a searcher, both for large single 
instances (Sect. \ref{cavity_method}) and in the thermodynamic limit of infinite 
system size (Sect. \ref{thdlim}). In Sect. \ref{results}, we present and 
discuss the results obtained, and close with a summary and concluding remarks in 
section \ref{conclusions}.

\section{Random Walk Framework} \label{RWM}
\subsection{Graphs, Random Search Strategies and their Analysis}

We will investigate the efficiency of search strategies using random graphs as search
spaces. A random graph $G$ is defined by a set $\mc{V}$ of vertices  and a set 
$\mc E$ of edges represented by an adjacency matrix $C =(c_{ij})$, with its entries 
$ c_{ij} $ taking the value 1 if nodes $i$ and $j$ are connected by an edge, and 
0 otherwise. We denote by $N = | \mc{V} |$ the number of vertices or nodes in the graph. 
We assume the networks to be undirected, so $c_{ij} = c_{ji}$ for each pair of nodes 
$(i,j)$, and that there are no self-loops in the system, hence $c_{ii}=0$ for all $i$.  

Our analysis of the efficiency of random search strategies will be based on recent work
of De Bacco et al. \cite{debacco2015average} who analyse the average number $S_i(n)$ of 
{\em different\/} sites visited by a random walker starting on vertex $i$ in a random
walk of $n$ steps, when $n$ becomes large. They express $S_i(n)$ as
\be
{S_i}{\lp n \rp} = \sum\limits_{j \in \mc{V}} {H}_{ij}(n)\ ,
\label{eqSi}
\ee 
where  $ H_{ij} \lp n \rp$ denotes the probability of visiting node $j$ at least once 
in the first $ n $ time steps when the walker started at node $i$. They evaluate the 
large $n$ asymptotics of this number in terms of its $z$-transform
\be
\hat S_i(z) = \sum_{n=0}^\infty S_i(n) z^n = \sum\limits_{j \in \mc{V}} \hat H_{ij}(z)\  .
\label{eqhSi}
\ee 
The $z$-transform $\hat H_{ij}(z)$ in turn is expressed in terms of the $z$-transform of
the $n$-step transition probability $G_{ij}(n) = (W^n)_{ij}$, with $W=(W_{ij})$ denoting
the matrix of probabilities for one-step transitions $i\to j$. De Bacco et al. 
\cite{debacco2015average} find
\be
\hat{H}_{ij}(z)=\frac{1}{1-z} \frac{\hat{G}_{ij}(z)}{\hat{G}_{jj}(z)}\ . 
\label{eqhHi}
\ee
In order to make this paper reasonably self-contained, we reproduce the key steps of this
derivation in Appendix \ref{app_RWF}. The large $n$ asymptotics of $S_i(n)$ is then extracted
by analysing the $z\to 1$ asymptotics of its $z$-transform. The  analysis in 
\cite{debacco2015average} covers the case where the random walker is performing an unbiased 
random walk, for which the probability of transitioning from node $i$ to node $j $ is given by 
\be
W_{ij} = \frac{c_{ij} }{k_i}\ ,
\label{eqWij}
\ee
where $ k_i =|\pr i|$ is the degree of node $i$,  with $\pr i$ denoting the set of the 
neighbours of node $i$.

Our analysis of the efficiency of random search strategies can fully utilise this theoretical
framework. It only requires two extensions. 

The first is minor. We mark a subset of vertices of the graph as having items of interest 
hidden on them. To do so, we associate an indicator variable $\xi_j $ with each site $j$ which 
designates whether an item is hidden on that node $(\xi_j =1)$, or not $(\xi_j =0)$.
Then
\be
{S_i}{\lp \bsl {\xi},n \rp} = \sum\limits_{j \in \mc{V}} {H}_{ij}(n) \xi_j 
\label{eqSixi}\ ,
\ee 
with $\bx = \{ \xi_1, \xi_2, ...., \xi_N \}$, will denote the average number of hidden items 
found in an $n$-step random walk starting at node $i$. Its large-$n$ asymptotics will again be 
analysed in terms of the $z\to 1$-asymptotics of its $z$-transform, which --- using Eqs. 
(\ref{eqhSi}) and (\ref{eqhHi}) --- we see to be given by
\be
\hat{S}_i(\bx, z)= \frac{1}{1-z} \sum\limits_{j \in \mc{V}} 
\frac{\hat{G}_{ij}(z)}{\hat{G}_{jj}(z)} \xi_j \ .
\label{eqhSixi}
\ee

The second modification is concerned with looking at a wider family of random walk models.
Rather than restricting the analysis to unbiased random walks, we will look at a large
family of degree-biased random walks with one step transition matrices given by
\be
W_{ij} = \frac{c_{ij} s(k_j)}{\gmm_i}\ .
\label{transitionprob}
\ee
Here $s(k)$ is a function of the degree, which we will refer to as a {\em search strategy}. The 
constant $ \gmm_i$ is dictated by the normalisation requirement for transition probabilities, 
giving
\be
\gmm_i= \sum\limits_{j} c_{ij} s(k_j) \ .
\ee

In order to evaluate (\ref{eqhSixi}), we need the $z$-transform of the matrix of $n$-step 
transition probabilities. It is given by
\be
\hat G(z) = \big[\bbbone - z W\big]^{-1}\ ,
\ee
with $\bbbone$ denoting the $N\times N$ identity matrix. The matrix $W$ of one-step transition
probabilities satisfies a detailed balance condition with the equilibrium distribution 
\be
p_i = \frac{1}{Y}\, \gmm_i s(k_i)\ ,\qquad \mbox{with}\qquad  
Y = \sum_{i \in \mc{V}} s(k_i)\gmm_i\ .
\label{eq_dist}
\ee
This can be used to express $\hat G(z)$ in terms of a symmetric matrix 
$\hat{R}(z)$ as
\be
\hat G(z) = D^{-\hf} \hat{R}(z) D^{\hf}\ ,
\label{GfromR}
\ee
where $D = \mbox{diag}(\gmm_i s(k_i)\big)$, and
 \bea
   \hat{R}(z) = \big [ \bbbone - z D^{\hf} W D^{-\hf} \big]^{-1} \ .
   \label{rhatfull} 
 \eea 
This matrix is easily seen to be a symmetric. The fact that $\hat G(z)$ and $\hat R(z)$
are related by a similarity transformation can be exploited \cite{debacco2015average} to analyse 
$\hat G(z)$ via the spectral decomposition of $\hat R(z)$. 
One has
\begin{eqnarray}
\hat{R}(z) &=&   \frac{ \vv_1  \vv_1^T}{1-z} + 
\sum\limits_{\nu=2}^{V} \frac{\vv_\nu \vv_\nu^T}{1-z \lambda_\nu} \equiv \frac{ \vv_1  \vv_1^T}{1-z} + \hat C(z)
\ ,
\label{Rhat}
\end{eqnarray}
where we have isolated the contribution of the Perron-Frobenius eigenvalue $\lambda_1=1$ of
$W$, and introduced $\hat C(z)$ to denote the contributions corresponding to the remaining
eigenvalues. The (normalised) Perron-Frobenius 
eigenvector $\vv_1$ has entries 
\be
v_{1,i} = \sqrt{p_i}= \sqrt{\frac{s(k_i)\gmm_i}{Y}}\ .
\label{firstevec}
\ee
Assuming that the graph $G$ is connected, we know that the Markov chain described by $W$
is irreducible, hence that the multiplicity of the largest eigenvalue is 1 and that $\lambda_\nu 
< 1$ for all $\nu \ne 1$ by the Perron-Frobenius theorem. In the $z\to 1$ limit, the contribution 
from the second term of the RHS of Eq. (\ref{Rhat}) is therefore negligible in comparison to 
the contribution from the first term. 

Following the reasoning of De Bacco et al. \cite{debacco2015average}, one can use this fact 
to determine the dominant $z\to 1$ asymptotics of $\hat{S}_i(\bx, z)$ in the large $N$ limit, 
and finds
\begin{eqnarray}
\hat{S}_i(\bx, z) &  \sim &  \frac{1}{(1-z)^2 Y} \sum\limits_{j \in \mc{V}} 
\frac{s(k_j)\gmm_j}{\hat{R}_{jj}}\,\xi_j\ ,\qquad z \to 1\ .  
\label{hatrdiv}
\end{eqnarray} 
In Eq. (\ref{hatrdiv}), the $N\to\infty$-limit is assumed to be taken, and we 
have introduced
\bea 
\hat{R}_{jj}  = \lim_{z \to 1}  \lim_{N \to \infty} \hat{R}_{jj}(z)\ . 
\label{rhatjj} 
\eea 
For the sake of completeness, the key steps of this derivation are reproduced in Appendix
\ref{app_SPA}.

Upon taking an inverse $z$-transform, the $1/{\lp 1-z \rp}^2$ divergence in Eq.
\eqref{hatrdiv} translates into a linear large-$n$  behaviour of $S_i(\bx, n)$
of the form
\be
S_i(\bx, n) \sim B\,n\ , \qquad  n\gg 1\ ,
\ee
with 
\be
B = \frac{1}{Y} \sum\limits_{j \in \mc{V}} \frac{s(k_j)\gmm_j}{\hat{R}_{jj}}\, \xi_j\ . 
\label{B}
\ee
Once more it is assumed that the $N\to\infty$-limit is taken in this expression. Note that
$S_i(\bx, n)$ is for large $n$ independent of the starting vertex $i$. We will in what follows 
refer to the constant $B$ as the {\em search efficiency}.

The non-trivial element in the evaluation of the search efficiency $B$ is related to the
$\hat{R}_{jj}$ that appear in the result, which according to Eq. (\ref{rhatfull}) are the 
diagonal elements of the inverse of a large matrix. We adopt the approach of 
\cite{debacco2015average} to evaluate these diagonal elements of inverse matrices in 
terms of single-site variances of a suitable multivariate Gaussian distribution, and use 
the cavity method to do this in practice for large systems. The method will be
explained below in Sect. \ref{sec_theoryB}. Before that, though, we turn to describing 
the hiding strategies that we consider in the present paper.

\subsection{Hiding Strategies} \label{hiding}
In order to be able to discuss efficiency of search strategies, we also need to specify the 
strategies according to which items are hidden in a network. We shall take these hiding
strategies to be probabilistic as well. One of the simplest choices is unbiased random 
hiding,  which can be characterised in terms of a Bernoulli distribution for the $\xi_j$ 
as
\be
p(\xi_j) = \rho_h\, \delta_{\xi_j,1} + (1-\rho_h)\, \delta_{\xi_j,0}
\ee 
for $0 < \rho_h < 1$. The parameter $\rho_h$ specifies the average fraction of vertices 
which have items hidden on them. As for search, we will look at a broader family of 
hiding strategies that are taken to be correlated with degree, i.e. we will choose
\begin{equation}
p \lp \xi_j = 1 | k_j=k \rp = \rho_h\,\frac{ h(k)}{ \la h \ra}\ , 
\label{Exi1ofk}
\end{equation} 
in which $h(k)$ is a function of the degree, which in what follows we will refer to as a 
{\em hiding strategy}, and  $\la h \ra = \sum_k p_k h(k)$. Note that for a given hiding 
strategy the range of achievable $\rho_h$ is bounded, as one needs to ensure that 
\be
 \max\limits_k\  \rho_h\,\frac{ h(k)}{ \la h \ra} \le 1\ .
\label{rhoh-limit}
\ee

In Table \ref{tab1} we list the families of degree biased hiding and search strategies 
that we will consider in the present paper, along with the parametrisations we use 
to explore each family. We have tried to cover the major functional forms for both 
hiding and search strategies. 

\begin{table}[h!]
\begin{center}
\begin{tabular}{ |l|c|c| }
\hline
Functional form & Hiding	& Searching  \\
\hline
power-law	& $h(k) = k^\bt$	& $s(k) = k^\al$\\
exponential	& $h(k) = \me^{\bt k}$	& $s(k) = \me^{\al k}$	\\
logarithmic	& $h(k) = \log(1 + \bt k^{\gm_h})$	& $s(k) = \log( 1 + \al k^{\gm_s})$ \\
\hline
\end{tabular}
\end{center}
\caption{Overview of the degree-biased hiding and search strategies and their paremetrisations
investigated in the present paper.}
\label{tab1}
\end{table}

We will investigate the effect of selecting a particular parameterised family of search 
strategies for any of the given hiding strategies and look at the dependence of the search
efficiency $B$ on the parameters characterising the search strategy. We will explore both 
matched and mismatched combinations of hiding and search strategies. Results will be presented 
in Sect. \ref{results}.

\section{Evaluation of Search Efficiencies} \label{sec_theoryB}
In this section we turn to the actual evaluation of search efficiencies. As mentioned
above, the non-trivial problem that must be solved in this evaluation is that of 
evaluating diagonal elements of the resolvent, i.e., diagonal element of an inverse
of a large matrix.

We will perform our analysis both for single large instances of the problem by
suitably adapting the cavity method developed for spectra of sparse symmetric random
matrices \cite{rogers2008cavity}, and in the thermodynamic limit by interpreting the
self-consistency equations for the inverse cavity variances arising from the cavity 
analysis as stochastic recursions. 

The single instance calculation, too, will very closely follow \cite{debacco2015average}, 
implementing the two extensions required to handle the search aspect and the more general 
(degree biased) random walk models we are looking at in the present paper. The transition
to the thermodynamic limit turns out to be non-trivial, though, and requires the introduction
of an infinite family of (degree dependent) distributions of inverse cavity variances.

Another theoretical problem that arises is related to the fact that, whereas the 
single-instance cavity analysis is performed on a simply connected network (in the case of 
random graph ensembles, on their giant component), the analysis of the thermodynamic limit, 
if naively performed, includes contributions from finite clusters of the system, which 
would have to give null-contributions to overall search efficiencies. We will utilise
methods recently devised in \cite{kuehn2016}, to obtain results for the thermodynamic
limit which are properly restricted to the giant component.

\subsection{Cavity Method} \label{cavity_method}

The evaluation of the $\hat{R}_{jj}$ appearing in Eq. (\ref{hatrdiv}) requires a 
matrix inversion \eqref{rhatfull}, followed by taking suitable limits \eqref{rhatjj}. 
The matrix inversion would be computationally expensive for large systems. It has
been pointed out in the context of evaluating spectra of random matrices 
\cite{EdwJon76} that expressing elements of inverse matrices as covariances of 
suitable multivariate Gaussians provides a simple method to compute inverse matrices,
which is particularly effective for large sparse systems \cite{rogers2008cavity}. It 
was used in \cite{debacco2015average} to evaluate the average number of sites visited 
by an unbiased random walker. 

For $z<1$ the matrix $\hat{R}(z)$ is positive definite, and so is its inverse. One
can therefore evaluate elements of $\hat{R}(z)$ as averages 
\be
\hat{R}_{ij}(z) = \langle x_i x_j \rangle
\ee
over the multivariate Gaussian
\begin{eqnarray}
P(\bsl {x}) & = & \frac{1}{Z}\,\exp  \lt[ - \hf {\bsl x}^T \hat{R}^{-1}(z) \bsl x \rt] 
\nn \\
 & = & \frac{1}{Z}\, \exp  \Bigg[ -\hf \sum\limits_{i,j\in  \mc{V} } {x_i} \bigg( \delta_{ij} - z 
c_{ij} \sqrt {\frac{{s(k_i) s(k_j)}  }{ {\gmm_i \gmm_j}}} \bigg) {x_j} \Bigg]\ .
\label{cv4}
\end{eqnarray}
To proceed it is advantageous \cite{kuehn2015matrix} to rescale variables $x_i/\sqrt{\gmm_i} 
\to x_i$. Keeping the same symbols for the rescaled variables, we have 
\bea
P(\bsl {x}) &=& \frac{1}{Z}\,\exp  \Bigg[ -\hf \sum\limits_{i,j\in  \mc{V} } {x_i} 
\bigg(\gmm_i\,\delta_{ij} - z\, c_{ij} \sqrt{s(k_i) s(k_j)} \bigg) {x_j} \Bigg] 
\nn \\
 & = & \frac{1}{Z}\, \exp  \Bigg[ -\hf \sum\limits_{i,j\in \mc{V}} c_{ij} \lp \hf \lp x_i^2 s(k_j)
 +x_j^2 s(k_i)\rp - z\,  \sqrt{s(k_i) s(k_j)} \, x_i x_j\rp 
 \Bigg]
\eea
for their joint distribution, where we have inserted the definition of the $\gmm_i$ in the 
second line and used the symmetry of the $c_{ij}$ to express the resulting distribution 
in terms of an exponential of a manifestly symmetric quadratic form.

To evaluate the diagonal elements $\hat{R}_{jj}$ in Eq. (\ref{hatrdiv}), one only
needs single-site variances $P(\bsl {x})$, for which only single-site marginals of
$P(\bsl {x})$ are needed. Following standard reasoning \cite{rogers2008cavity,
kuehn2015matrix, debacco2015average} one finds these as
$$
P_i( {x_i} ) 
=  \int  \Big[ \prod_{j \in \mc{V} \backslash i} \dd x_j \Big]  P(\bsl x)\ ,
$$
giving
\be
P_i( {x_i} ) \prp \int \Big[ \prod\limits_{j \in \pr i}  \dd x_j \Big]   
\hspace{3 pt} \text{exp} \Bigg[ -\sum\limits_{j \in  \pr i } \lp \hf\lp x_i^2 s(k_j) + x_j^2 s(k_i) \rp -  
 z \sqrt{s(k_i) s(k_j)}\, x_i x_j \rp \Bigg] \hspace{3 pt} 
P^{(i)}(\bsl x_{\pr i})\ .
\label{marg}
\ee
Here $P^{(i)}(\bsl x_{\pr i})$ is the joint cavity marginal of the $x_j$ on sites $j$ which are 
neighbours of node $i$ on the graph with node $i$ missing. On a locally tree-like graph one 
has $P^{(i)}(\bsl x_{\pr i}) \simeq \prod\limits_{j \in \pr i}  \lp  P_j^{(i)}(x_j) \rp$,  
where $\pr i$ denotes the set of neighbours of $i$ and the  $P_j^{(i)}(x_j)$ are the 
single-site cavity marginals of $x_j$. Hence the integrals in Eq. (\ref{marg}) factor, and
\begin{eqnarray}
P_i( {x_i} ) & \prp & \prod\limits_{j \in \pr_i} \int  \dd x_j \
\text{exp} \lt[ -\hf \lp x_i^2 s(k_j) + x_j^2 s(k_i) \rp + z\, 
\sqrt{s(k_i) s(k_j)}\,x_i x_j \rt]  P_j^{(i)}(x_j)\ .
\label{eq13}
\end{eqnarray}
Following the same line of reasoning for single-site cavity marginals $ P_j^{(i)}(x_j) $, we have 
\be
P_j^{(i)}(x_j) \prp \prod\limits_{\ell \in \pr j \backslash i}  \int   \dd 
x_\ell  \ \text{exp}         
\lt[ -  \frac{1}{2}     \lp x_\ell^2 s(k_j) + x_j^2 s(k_\ell) \rp + z 
  \sqrt{s(k_j) s(k_\ell)}\ x_j x_\ell \rt]\ P_{\ell}^{(j)}(x_\ell) \ .    
\label{eq73}
\ee    
The system \eqref{eq73} of equations is self-consistently solved by Gaussians of the form
\be
P_j^{(i)}(x_j) = \sqrt{\frac{\om_j^{(i)}}{2\pi}}\ \exp\lt[-\hf \om_j^{(i)} x_j^2\rt]\ ,
\ee
with $\om_j^{(i)} > 0$, entailing that the inverse cavity variances need (in the limit $z \to$ 1)
to satisfy the self-consistency equations
\be
\om_j^{(i)} = { \sum\limits_{ \ell \in \pr_j \backslash i} \lb s(k_{\ell}) - 
\frac{s(k_j) s(k_{\ell})}{{\om_{\ell}^{(j)}}+s(k_j)} \rb }\ . \label{eq61a} 
\ee
With the $P_j^{(i)}(x_j)$ Gaussian, the single site marginals $P_i(x_i)$ are also 
Gaussian. Denoting inverse single-site variances by $\om_i$, we obtain these in
terms of inverse cavity variances as
\be
\om_i = {{\sum\limits_{j \in \pr_i} \lb s(k_j) - \frac{s(k_i) 
s(k_j)}{{\om_j^{(i)}}+s(k_i)} \rb}}\ . \label{eq61b} 
\ee
Eqs. (\ref{eq61a}) and (\ref{eq61b}) generalise those obtained in \cite{debacco2015average} 
to cover general degree-biased random walk models.

Once Eqs. (\ref{eq61a}) are solved for a given single instance of a graph, the inverse 
single-site marginals can be computed. When evaluating the search efficiency $B$ according 
to Eq. (\ref{B}) we need to recall that the $\om_j$ are inverse single site variances of 
{\em rescaled\/} variables $x_j/\sqrt{\gmm_j}$. We therefore have $\hat R_{jj} = \gmm_j/\om_j$,
hence
\be
B =  \frac{1}{Y}\sum\limits_{j \in \mc{V}} s(k_j)\,\om_j\,\xi_j\ .
\label{eq262}
\ee 
We will see in Sect. 4 that, for sufficiently large systems, results obtained using the present 
approach agree very well with those of simulations. Before turning to results, however, we
will first elaborate the theory for the limit of infinitely large systems.

\subsection{Thermodynamic Limit} 
\label{thdlim}
In the thermodynamic limit, Eqs. \eqref{eq61a} can be interpreted as stochastic
recursion relations for inverse variances of single-site cavity marginals. In what
follows we will use these equations to obtain a system of self-consistency equations 
for the {\em distributions\/} of the inverse cavity variances for ensembles of random 
graphs in the configuration model class. It turns  out that degree dependent families 
of such distributions are needed due to the node degrees appearing in Eqs. \eqref{eq61a}. 
The resulting self-consistency equations for these distributions can be solved by a 
stochastic population dynamics algorithm \cite{mezard2001bethe}. The solution then 
determines the degree dependent distributions of inverse single site marginals needed 
to evaluate search efficiencies in the thermodynamic limit according to Eq. (\ref{eq262}).

However, it turns out that the results of this approach cannot be directly compared to single 
large instance calculations or to simulations, which are usually performed by restricting 
attention to the (single) giant component of a graph, whereas standard random graph 
ensembles typically describe systems which --- apart from the giant component --- also 
contain finite clusters. In Sect. \ref{sec_gc} below we will therefore introduce the 
necessary modifications which will allow one to compute search efficiencies of random 
walkers restricted to the giant component of random graph ensembles.

\subsubsection{Distributions of Inverse Cavity Variances}
The self-consistency equations \eqref{eq61a} for the inverse cavity variances imply that
$\om_j^{(i)}$ depends on the node degree $k_j$ of node $j$. Analogous degree dependences 
must therefore be expected for the $\om_\ell ^{(j)}$ appearing on the r.h.s. of  Eq.\eqref{eq61a}.  
In the thermodynamic limit, we therefore need to self-consistently determine an {\em entire 
family $\{\tilde{\pi}_k(\tilde{{\om}})\}_{k\ge 1}$ of degree-dependent distributions\/} of 
inverse cavity variances $\tilde{\om}$. 

Suppose that $k_j=k$. The probability $\tilde{\pi}_k (\tilde\om)\,\dd\tilde \om$ that 
$\om_j^{(i)} \in (\tilde{\om},\tilde{\om}+\dd \tilde{\om}]$ is obtained by summing over the 
probabilities of all realisations of the r.h.s. of \eqref{eq61a} which give a value in that 
range. Recall that $\om_j^{(i)}$ has contributions from all vertices adjacent to $j$, except 
$i$. Denoting by $\{k_\nu\}_{k-1} = \{k_\nu; \nu=1,\dots, k-1\}$ the set of degrees of 
the $k-1$ vertices adjacent to $j$ (not including $i$) which appear on the r.h.s. of 
Eq.\eqref{eq61a}, we have
\be
 \tilde{\pi}_k \lp \tilde{\om} \rp =    \sum\limits_{ \{ k_{\nu} \geq 1 \}_{k-1} 
}  \Big[ \prod_{\nu=1}^{k-1} \frac{k_{\nu}}{c} p_{k_{\nu}} \Big] \int 
\Big[ \prod_{\nu=1}^{k-1}   \text{d}{\tilde{\pi}}_{k_{\nu}}(\tilde{\om}_{\nu}) \Big]
\hspace{3 pt}  \delta \lt[ \tilde{\om} -  \Omega_{k-1} \lp \{ {\tilde{\omega}}_{\nu}, 
k_{\nu} \}| k \rp  \rt]\ .   
\label{eq35a}
 \ee
Here $p_k$ denotes the probability of having a vertex of degree $k$ in the graph, so
that $\frac{k}{c} p_k$ is the probability that a randomly chosen neighbour of a node
has degree $k$, with $c=\langle k\rangle$ denoting the mean degree. We have also 
introduced 
\be 
\Omega_q \lp \{ {\tilde{\omega}}_{\nu}, k_{\nu} \}| k \rp = 
\sum\limits_{\nu=1}^{q} \lp s \lp k_\nu \rp - \frac{s(k) s(k_{\nu}) 
}{\tilde{\omega}_{\nu} + s(k)} \rp\ , \label{Omg2}
\ee
and we have used the shorthand notation $\text{d}{\tilde{\pi}}_{k_{\nu}}(\tilde{\om}_{\nu}) =  
{\tilde{\pi}}_{k_{\nu}}(\tilde{\om}_{\nu}) \dd \tilde\om_\nu$. 

In a similar vein, the degree dependent distributions $\pi_k(\om)$ of inverse single-site
variances of the rescaled Gaussian variables $x_j$ are obtained from Eq. \eqref{eq61b} 
as
\begin{eqnarray}
 {\pi}_k \lp \om \rp =    \sum\limits_{ \{ k_{\nu} \geq 1 \}_{k} }   \Big[ 
\prod_{\nu=1}^{k} \frac{k_{\nu}}{c} p_{k_{\nu}} \Big] \int \Big[\prod_{\nu=1}^{k}   
\text{d}{\tilde{\pi}}_{k_{\nu}}(\tilde{\om}_{\nu})\Big] \  
\delta \lt[ {\om} -\Omega_k \lp \{ {\tilde{\omega}}_{\nu}, k_{\nu} \}| k \rp \rt]\ .   
\label{eq35b}
\end{eqnarray}
They can be evaluated once the solutions $\{\tilde \pi_k(\tilde\om)\}$ of Eq. 
(\ref{eq35a}) have been found. These distributions can be used to compute the search 
efficiency $B$ from Eq. (\ref{eq262}).

\subsubsection{Search Efficiencies} \label{thdlimB}
To evaluate search efficiencies, we rewrite Eq. (\ref{eq262}) as
\be
B =  \frac{1}{Y/N}\Bigg[ \frac{1}{N}\sum\limits_{j \in \mc{V}} {s(k_j) }{\om_j}\,\xi_j\Bigg] \ , 
\label{eq262n}
\ee 
thereby highlighting the fact that it is a ratio of two terms, which --- in the 
thermodynamic limit $N\to\infty$ --- can both be evaluated by appeal to the 
law of large numbers. Recalling from Eq. (\ref{firstevec}) that the normalisation constant 
$Y$ appearing in the Perron-Frobenius eigenvector $\vv_1$ of $\hat R$ gives
$$
\frac{Y}{N} = \frac{1}{N} \sum_{j \in \mc{V}} {s(k_j)} \gmm_j\ ,
$$
we find that this results in
\be
B = \frac{1}{\mathcal  N}\ \sum_k p_k\, \Big[ s(k)\ \mathbb{E}[\om|k]\ \mathbb{E}[\xi|k] \Big]
\label{Blim}
\ee
as the limiting expression for the search efficiency. Here
\bea
\mathbb{E}[\om|k] &=& \int {\rm d} \pi_k (\om) \, \om \nn\\
& = & \sum\limits_{\{ k_{\nu}\geq 1 \}_k}  
\Big[\prod\limits_{\nu=1}^k \frac{k_{\nu} }{c} p_{k_{\nu}}\Big]\ 
\int \Big[\prod\limits_{\nu=1}^{k} d \tilde{\pi}_k (\tilde{\om}_\nu)\Big] \hspace{2 pt}  
\Omega_k \lp \{ {\tilde{\omega}}_{\nu}, k_{\nu} \}| k \rp
\label{avomk}
\eea
by Eq. (\ref{eq35b}), and we have
\be
\mathbb{E}[\xi|k] = \rho_h\,\frac{ h(k)}{ \la h \ra}
\ee
from Eq. (\ref{Exi1ofk}), while 
\be
\mathcal N = c \Big[\sum_k \frac{k}{c} p_k s(k)\Big]^2
\label{cN}
\ee
is the limiting value of the normalisation factor $Y/N$. We refer 
to Appendix C for its evaluation.

The search efficiency $B$ clearly has a natural decomposition in terms of contributions 
of vertices of different degree. It can be written as 
\begin{eqnarray}
B = \sum\limits_{k \ge 1} p_k B_k\ ,
\end{eqnarray}
where the $k$-dependent components $B_k$ are given by
\begin{eqnarray}
B_k = \frac{1}{\mathcal N}\, s(k)\, \mathbb{E}[\om|k]\ \mathbb{E}[\xi|k]   
 \ .
\label{Bk_search_n}
\end{eqnarray}
They denote the fraction of sites of degree $k$ on which items are found per unit time in 
the course of an $n$-step degree-biased random walk.

The reasoning in the present section does not properly take into account the fact
that simulations or single-instance cavity analyses are typically performed on graphs 
which consist of a {\em single component}, given that a random walker can only explore the 
graph component on which (s)he starts the search in the first place. If that component
is one of the finite clusters, then only that finite cluster can be explored in the 
search so that the number of items found in a random walk will be finite, hence the 
efficiency of the search as defined by the number of items found per unit time in 
an $n$-step walk will tend to zero in the large-$n$ limit. In the following section we
will discuss the modifications of the theory necessary to take into account the fact
that only random searches on the giant component of a random graph will give a non-zero
contribution to the search efficiency $B$.

\subsubsection{Isolating Giant Component Contributions} \label{sec_gc}
As we have just indicated, any node belonging to one of the finite clusters 
would give a zero contribution to the search efficiency $B$ in the thermodynamic 
limit, and only the nodes in the giant component are going to contribute to the result.
It is therefore important to differentiate between the two and to be able to restrict 
results obtained for the search efficiency in the thermodynamic limit to contributions 
coming only from the giant component of the system. 

In order to do this, we can follow \cite{kuehn2016}, and supplement the recursions 
Eq. (\ref{eq61a}) for the inverse cavity variances and expression Eq. (\ref{eq61b}) 
for inverse single-site variances by analogous equations describing whether or not a 
site adjacent to a cavity belongs to the giant component of the system, and similarly 
whether a randomly selected site does or does not belong to the giant component.

This is achieved by introducing indicator variables $n_i$ for each node $i$ 
which take the value 1, if node $i$ belongs to the giant component of a graph and
0, if it doesn't. In a similar vein, indicator variables $n_j^{(i)}$ are introduced
to express whether a node $j$ adjacent to a cavity site $i$ does or does not belong
to the giant component. For these we have
\bea 
n_i &=& 1 - \prod_{j\in\partial i}\big(1- n_j^{(i)}\big)
\label{ind}\\
n_j^{(i)} &=& 1 - \prod_{\ell\in\partial j\setminus i} \big(1- n_\ell^{(j)}\big)\ .
\label{cavn}
\eea
The first of these equations states that node $i$ belongs to the giant component of 
the graph if at least one of its neighbours is connected to the giant-component through
a path not involving $i$, whereas the second equation expresses the same fact for 
a site adjacent to the node $i$ on the cavity graph from which node $i$ is removed.

In the thermodynamic limit Eqs. (\ref{cavn}) can once more be thought of as stochastic
recursions for random cavity indicator variables. For a node $j$ of degree $k_j=k$ 
adjacent to a cavity node $i$ we now seek to determine the joint probability 
$\tilde\pi_k (\tilde\om, \tilde n) \,\dd\tilde \om$ that the inverse cavity variance
$\om_j^{(i)}$ falls into the infinitesimal interval $(\tilde \om,
\tilde{\om}+\dd \tilde \om]$ {\em and\/} that the cavity indicator variable $n_j^{(i)}$
takes the value $n_j^{(i)}= \tilde n \in \{0,1\}$. As for Eq. \eqref{eq35a}, this joint 
probability is obtained by summing over the probabilities of all realisations of the r.h.s. 
of Eqs. \eqref{eq61a} and \eqref{cavn} which give a value of the inverse cavity variance 
in that prescribed range {\em and\/} a value $\tilde n$ for the cavity indicator 
variable. This gives
\begin{eqnarray}
  \tilde{\pi}_k \lp \tilde{\om}, \tilde{n} \rp &=&    \sum\limits_{ \{ k_{\nu} 
\geq 1, \tilde{n}_\nu \}_{k-1} } \Big[\prod_{\nu=1}^{k-1} \frac{k_{\nu}}{c} 
p_{k_{\nu}}\Big]  \int \Big[\prod_{\nu=1}^{k-1}   
\text{d}{\tilde{\pi}}_{k_{\nu}}(\tilde{\om}_{\nu},\tilde{n}_\nu)\Big] \hspace{3 pt}  
\delta \lt[  \tilde{\om} - \Omega_{k-1} \lp \{ {\tilde{\omega}}_{\nu}, k_{\nu} \}| k \rp
\rt] \nn \\
  & &  \hspace{15 pt} \times  \delta_{ 
\tilde{n},1-\prod_{\nu=1}^{k-1} \lp 1 - \tilde{n}_{\nu} \rp} \ .
\label{eq37a} 
  \end{eqnarray}
In a similar vein we obtain the joint distribution $\pi_k (\om, n)$ for the inverse 
single-site variances $\om_i$ and the single-site indicator variables $n_i$ 
from the solution of Eq. \eqref{eq37a} as
  \begin{eqnarray}
 {\pi}_k \lp \om, n \rp &=&    \sum\limits_{ \{ k_{\nu} \geq 1, \tilde{n}_\nu 
\}_{k} }  \Big[\prod_{\nu=1}^{k} \frac{k_{\nu}}{c} p_{k_{\nu}}\Big]  \int 
\prod_{\nu=1}^{k}   
\text{d}{\tilde{\pi}}_{k_{\nu}}(\tilde{\om}_{\nu},\tilde{n}_{\nu}) \  \delta 
\lt[  {\om} - \Omega_{k} \lp \{ {\tilde{\omega}}_{\nu}, k_{\nu} \}| k \rp
\rt]  \nn \\ 
 & & \hspace{15 pt} \times \delta_{n,1-\prod_{\nu=1}^{k} \lp 1 - 
\tilde{n}_{\nu} \rp} . 
\label{eq37b}
\end{eqnarray}

The search efficiency $B$ evaluated on the giant cluster can be written as 
\be
B =  \frac{1}{Y_g/N_g} \Bigg[ \frac{1}{N_g} \sum_{j \in \mc{V}_g} {s(k_j)\, 
}{\om_j}\,\xi_j \Bigg]\ , \label{eq263}
\ee 
where $\mc{V}_g$ is the set of nodes in the giant cluster, $N_g$ is the number 
of nodes in the giant cluster and $Y_g = \sum\limits_{j \in \mc{V}_g} {s(k_j) 
}{\gmm_j}$. In the thermodynamic limit, both the numerator and the denominator 
in this expression are once more evaluated by appeal to the law of large numbers.
We will use a recent result of Tishby et al. \cite{Tishby+18} about degree distributions 
conditioned on the giant component of random graphs to evaluate the denominator 
and use the $\pi_k(\omega,n)$ in \eqref{eq37b} to compute conditional expectations 
of inverse single-site variances $\omega$ conditioned on degree {\em and\/} on 
nodes belonging to the giant cluster
\bea
\mathbb{E}[\om|k,n=1] &=& \int {\rm d} \pi_k (\om|1) \, \om \nn\\
& = & \frac{1}{\rho}\sum\limits_{\{ k_{\nu}\geq 1, \tilde{n}_\nu\}_k} 
\Big[\prod\limits_{\nu=1}^k \frac{k_{\nu} }{c} p_{k_{\nu}}\Big]\ 
\int \Big[\prod\limits_{\nu=1}^{k} \dd \tilde{\pi}_k (\tilde{\om}_\nu,\tilde n_\nu)\Big] \,  
\Omega_k \lp \{ {\tilde{\omega}}_{\nu}, k_{\nu} \}| k \rp \nn\\
& &~~~~~~~~~~~~~~~~~~~~~~~~~~~~~~\times \delta_{1,1-\prod_{\nu=1}^{k} \lp 1 - 
\tilde{n}_{\nu} \rp}\ ,
\label{avomkn1}
\eea
in which $\rho$ is probability of a randomly chosen vertex to belong to the giant cluster. 
This gives
\be
B = \frac{1}{\mathcal N_g}\sum_{k \geq 1} p(k|1) \,\Big[ s(k)\,\mathbb{E}[\om|k,n=1]\,
\mathbb{E}(\xi | k)\Big]
 \ ,
\label{Blimg}
\ee
with $p(k|1)$ denoting the degree distribution {\em conditioned\/} on the giant cluster
\cite{Tishby+18}, and
\be
\mathcal N_g = \frac{c}{\rho}\ \sum\limits_{k,k^{\prime}} \frac{k }{c}p_k\   
\frac{k^{\prime} }{c} p_k^{\prime}\, s(k) s(k^{\prime}) \big[ 
{1-{(1-\tilde{\rho} )}^{{k^\prm} +k -2}  }\big] 
\label{gcnorm}
\ee
giving the limiting value of $Y_g/N_g$; its evaluation, following \cite{Tishby+18}, is left 
to Appendix C. In Eq. \eqref{gcnorm}, $\tilde{\rho}$ denotes the probability of a random link 
pointing to nodes on the giant cluster. These quantities can be easily evaluated using standard generating 
function techniques.

As before, the expression for the search efficiency restricted to the giant component
has a natural decomposition into contributions of vertices of different degrees. In the 
present case, we have
\begin{eqnarray}
B_k = \frac{1}{\mathcal N_g}\,s(k)\,\mathbb{E}\big[\om|k,n=1\big]\,\mathbb{E}\big[\xi | k\big]\ .
\label{Bpk_search_n}
\end{eqnarray}

The self consistency equations (\ref{eq37a}) for the $\tilde\pi_k(\tilde\omega,\tilde n)$ which are 
needed to evaluate search efficiencies in the thermodynamic limit are very efficiently solved using a 
stochastic population dynamics algorithm. The new aspect in the present problem is that several 
such populations are needed to represent the $\tilde\pi_k(\tilde\omega,\tilde n)$ for the 
different degrees $k$ in the system.

\subsection{Analytic Results for Random Regular Graphs}
On random regular graphs, we have $p_k = \delta_{k,c}$. Hence there cannot be a non-trivial
degree biased strategy, as the normalised matrix of transition probabilities is independent of
the choice of $s(k)=s(c)=s$, and we are therefore looking at an unbiased random 
walk as the search strategy, and random hiding as the hiding strategy.

Given that all nodes (and all links) are equivalent in the thermodynamic limit, the solution
of Eq. \eqref{eq35a}, \eqref{Omg2} is $\tilde\pi_c(\tilde \om) = \delta(\tilde\om -\bar\om)$,
with $\bar\om$ satisfying
\be
\bar{\om} = (c-1) \lb s - \frac{s^2}{\bar{\om} + s} \rb\ .
\label{rr1}
\ee 
The only non-zero solution to \eqref{rr1} is $\bar{\om} = s (c-2)$, where $s=s(c)$. Using
$\mathcal N = c s^2$ from Eq. (\ref{cN}), and inserting $\tilde\pi_c(\tilde \om) = 
\delta(\om -\bar\om)$ into Eq. \eqref{eq37b} we have
\be
\mathbb E[\om|k] =\mathbb E[\om|c]  = cs \frac{c-2}{c-1} \ ,
\ee
and thus
\begin{eqnarray}
B = \rho_h \frac{c-2}{c-1}\ .
\label{rrB2}
\end{eqnarray} 
This result is independent of $s$ as it should. The result was obtained in 
\cite{debacco2015average} from a single-instance cavity analysis of the case 
$\rho_h=1$.
\subsection{Approximations}
\label{approx}
In what follows we will look at two approximate descriptions of the hide and seek
problem. 

The first is based on comparing the equilibrium distribution of the random
walker executing the search with the distribution characterising the location of 
hidden items on the network. While this equilibrium type analysis does not actually
provide us with an estimate of the search efficiency, it will allow us to find 
parameter settings for the strategy of the searcher which will optimise the search
efficiency for a given hiding strategy.

The second approximation is based on a so-called non-backtracking assumption and it
will actually produce approximate values for search efficiencies. For reasons to be
described below, we expect these approximations to become quite accurate in the limit
where most vertices of the system actually have large degrees.

\subsubsection{An Analysis Using Equilibrium Distributions}
The analysis in the present section is based on the observation that a random walker
starting her walk on any randomly chosen site of a network will --- after only a few 
steps of the walk --- very quickly ``forget" about any specific properties of the 
starting vertex and start visiting different vertices of the system with probabilities
given by the equilibrium probability of the random walker.

Let us denote by $q_s(k)$ the equilibrium probability of the random walker to visit a site 
of degree $k$, and by $q_h(k)$ the probability that a randomly selected site {\em with\/} an 
item hidden on it has degree $k$. Choosing parameters of the search strategy in such a way 
that $q_s$ is as close as possible to  $q_h$ should then provide a good heuristic to optimise 
the efficiency of a search strategy.

From Eq. \eqref{Exi1ofk} we have
\be
q_h(k) = p_k \frac{h(k)}{\la h \ra}\ ,
\ee
with $\la h \ra= \sum_k p_k h(k)$ for the conditional probability that a site has 
degree $k$ given that an item is hidden on it. In a similar fashion we have 
\be
q_s(k) = \sum_{i\in \mc V} p_i \delta_{k_i,k} = p_k \frac{k s(k)}{\la k s\ra}
\ee
for the probability that a random walker in equilibrium finds herself on a site 
of degree $k$.

A measure of the similarity of $q_s$ and $q_h$ is given by the Kullback-Leibler (KL) 
divergence between them, which is given by 
\begin{equation}
\text{KL}({q}_s||q_h) = \sum\limits_{k=1}^{\infty} {q}_s(k)\, {\text{log}}\,\lb 
\frac{{q}_s(k)}{q_h(k)} \rb = \sum_{k=1}^{\infty} \frac{k s(k)}{\langle k s\rangle} p_k\,
\log \lb \frac{k s(k)}{\langle k s\rangle} \frac{\langle h\rangle}{h(k)} \rb\ .
\label{Kl2}
\end{equation}
Minimising the KL divergence over any parameters characterising the search strategy is 
then expected to provide a good indication of the parameter setting for the most efficient 
search strategy within the parameterised family of strategies under consideration.

For power-law search $s(k)=k^\alpha$ pitted against the power-law hiding strategy 
$h(k)=k^\beta$, the minimisation of the KL divergence can be done analytically, and it
leads to
\be
\alpha= \beta - 1
\ee
for the exponent of the most efficient search strategy. We will see in the results
section that the result is far from exact. The main reason is, of course, that the 
number of marked sites visited at least once is what matters for the search 
efficiency whereas the total frequency of visits (including repeated visits) to 
sites is the quantity determining the equilibrium distribution.

\subsubsection{A Non-Backtracking Approximation}
On a network in which degrees are typically large, the probability of an unbiased random 
walker to return to the site from which she transitioned to the site she is currently on 
becomes {\em small\/}. This is because the probability of choosing any particular neighbour 
as the target of the next step, and thereby the probability of retracing the last step is 
inversely proportional to the degree of the site the random walker currently finds herself 
on, which is therefore small for a site with a large number of neighbours. 

For a degree-biased random walker, this effect will persist unless the degree-bias in the 
transition probabilities is extremely strong (e.g. such that the walker almosr always goes 
to the neighbouring site with the largest degree: one can easily convince oneself that in 
such a situation there can be configurations of neighbouring sites at which the random
walker could be trapped once it hits such a set of sites.).

Thus, assuming that back-tracking events are rare, a non-backtracking approximation can 
be formulated as follows. Denote by $S(\bm\xi,n)$ the average number of items found in an 
$n$-step walk. We suppress the index of site from which the walker started out its search, 
as we have learned above that for large $n$ the search efficiency will be independent of the 
starting site. Assume that the random walker at the $n$-th step of her walk visits the 
site $j$, coming from a site $i$, which is adjacent to $j$. If site $j$ is visited for 
the first time, the only chance to find additional items in the next step is {\em not 
to backtrack\/} on the previous step, but to visit sites $\ell\in\partial j\setminus i$. 
Taking averages over the sites, using the equilibrium distribution $p_i$ to give the 
probability that the walker found herself on site $i$ in step $n-1$, and assuming that
$j$ and its neighbours (apart from $i$) are being visited for the first time, we obtain
\be
S(\bm\xi,n+1) = S(\bm\xi,n) + \sum_{i\in\mc V} p_i \sum_{j\in\partial i} W_{ij}
\sum_{\ell\in\partial j\setminus i} W_{j\ell}\, \xi_\ell\ .
\ee
This recursion is easily solved. Taking $S(\bm\xi,0)=0$ as the initial condition, we get
\be
S(\bm\xi,n) = B\, n\ ,
\ee
with
\be
B = \sum_{i\in\mc V}\ p_i \sum_{j\in\partial i} W_{ij} \sum_{\ell\in\partial j\setminus i} 
W_{j\ell}\xi_\ell\ .
\ee
Using the transition probabilities \eqref{transitionprob} and the resulting expression 
\eqref{eq_dist} for the equilibrium distribution we obtain
\be
B=\frac{1}{Y} \sum_{i\in\mc V} s(k_i)\gmm_i\ \sum_{j\in\partial i}\frac{s(k_j)}{\gmm_i}
\sum_{\ell\in\partial j\setminus i} \frac{s(k_\ell)}{\gmm_j} \xi_\ell\ .
\label{Bnonbtr}
\ee
Repeating the line of reasoning that lead to the expression \eqref{Blim} for search efficiencies
in the thermodynamic limit, we can evaluate Eq. \eqref{Bnonbtr} in this limit.
The resulting expression is
\be
B = \frac{c}{\mc N} \sum_k \frac{k}{c} p_k s(k) \sum_{k'} \frac{k'}{c} p_{k'} s(k')
\sum_{\{k_\nu\}_{k'-1}} \Big[ \prod_{\nu=1}^{k'-1} \frac{k_\nu}{c} p_{k_\nu} s(k_\nu)\Big] 
\sum_{\nu=1}^{k'-1} \frac{s(k_\nu)\,\mathbb E[\xi|k_\nu]}{s(k) +\sum_{\nu=1}^{k'-1} s(k_\nu)}\ .
\ee
We shall see in Sect. \ref {secHandS} below that this approximation is remarkably efficient 
even for systems with moderate values of their mean degree.

\section{Results} \label{results} 
We now turn to results. We will evaluate search efficiencies for large finite systems 
using {\bf (i)} the single-instance cavity approach described in Sect. \ref{cavity_method} 
and {\bf (ii)} numerical simulations. Search efficiencies in the thermodynamic limit will 
be analysed using {\bf (iii)} the methods described in Sect. \ref{thdlim}. We will use 
these three approaches to explore how various search strategies fare against a range of 
hiding strategies. We shall find that there is an excellent agreement between results 
obtained using simulations and those obtained using either the single-instance cavity 
method or the method designed for the large system limit, provided the projection to the 
giant component described in Sect. \ref{sec_gc} is used in the thermodynamic limit, and
finite single instances are sufficiently large.

As mentioned in Sect. \ref{hiding}, we cover several functional forms for the degree
bias of both hiding and searching strategies, namely, power-law, exponential, and 
logarithmic strategies. We evaluate search efficiencies across the spectrum of 
functional forms used to describe hiding and searching strategies and --- for given 
functional forms --- across parameter ranges characterising them. 

Finally we investigate search efficiencies for various graph types, including
\er graphs and scale-free graphs, and we assess the quality of our approximate 
approaches by comparing them with exact results.

\subsection{Validating the Theory}
We begin by validating the theoretical approaches described in Sect. \ref{sec_theoryB},
by comparing their results with those of stochastic simulations. We do this initially
for complete occupancy $\xi_i\equiv 1$, where the number of different sites visited
by a walker is a measure of network exploration efficiency (rather than search 
efficiency). 

Random networks of a sufficiently large size are generated, and $n$-step degree biased 
random walks starting from a randomly chosen vertex on the giant cluster are simulated. 
The number of different sites visited is recorded. As noted in Sect. \ref{sec_theoryB} 
that number is for sufficiently large $n$ expected to be independent of the starting 
vertex and $S(n) \sim Bn$ for $1 \ll n \ll N$. We determine the exploration efficiency
$B$ by averaging over many realisations of the random walk and over many realisations 
of random graphs in the given ensemble. Alternatively, we compute the exploration 
efficiency $B$ directly using the cavity method, averaging the results over the same
set of graphs. The cavity method requires to solve Eqs. \eqref{eq61a} for single large
instances. We found that for all cases considered in the present paper, this can very
effectively be done by simple forward iteration.

\begin{figure}[h]
\centering
${\includegraphics[width=0.475\textwidth]{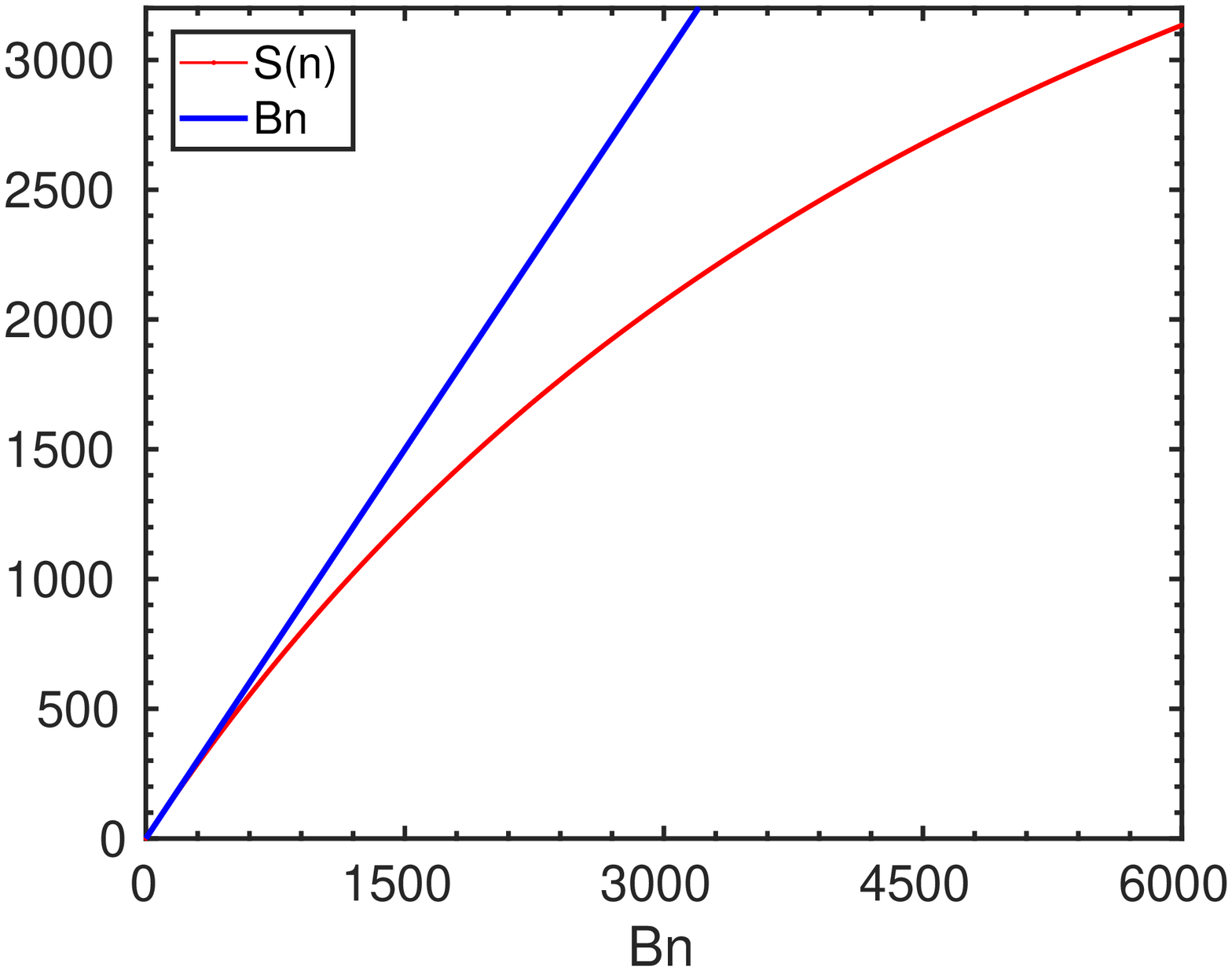}\atop ~}$
\hfil
\includegraphics[width=0.455\textwidth]{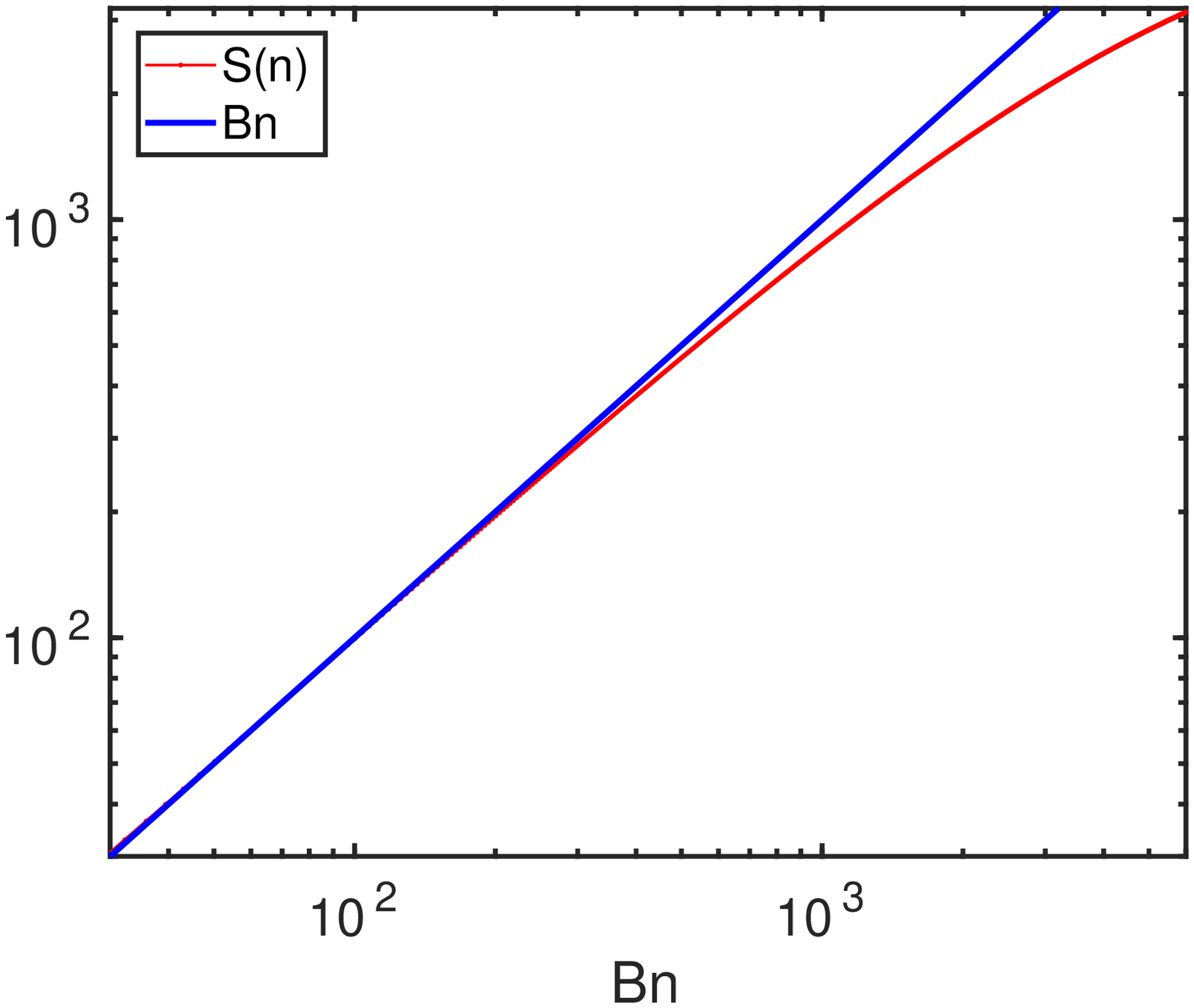}
\caption{Behaviour of $S(n)$ for a degree-biased random walker with degree bias following a 
power-law $s(k)=k^\alpha$ with $\alpha=1$. The left panel displays both $S(n)$ (lower curve)
and $Bn$ (upper curve), with $B\simeq 0.716872$ determined from simulations, as functions of $Bn$.
The right panel shows the same results on a double-logarithmic plot. The behaviour of $S(n)$ 
is well described by the linear law for not too large $n$. For larger $n$ there is a clear 
crossover to sub-linear behaviour due to finite size effects. Results were obtained for \er
graphs of mean degree $c$=4. Simulations were performed on the giant component of graphs 
whose original size was $N=6000$. For $c=4$ the giant component occupies a fraction $\rho\simeq 
0.98$ of the entire system. Results of simulation runs are averaged over $N_s = 2000$ random 
graph realisations.} \label{Svsn1}
\end{figure}

For finite single instances we find that graph sizes $N=6000$ were sufficient to compare
simulation results with those obtained via the cavity method on the one hand side, and
with the thermodynamic limit results on the other hand side. All finite single instance
results shown below will therefore have been obtained for systems of this size. The 
optimal $n$ range for which the behaviour of $S(n)$ can be fitted by a linear law has
been determined by minimising $\chi^2$ in linear regression. Fig. \ref{Svsn1} shows 
the results of simulations and confirms the linear behaviour of $S(n)$ for intermediate 
$n$.  We found that $40\lesssim n \lesssim 230$ was an optimal range for the linear
fit for this system, but observed that slightly narrower fitting ranges were required 
for other degree biases. From the simulations we determine $B= 0.716727 \pm 0.000203$. This 
compares well with the analysis of $B$ evaluated directly via the cavity approach, 
which gives $B= 0.716789 \pm 0.000210$.\\

\begin{figure}[t]
\centering
\includegraphics[width=0.475\textwidth]{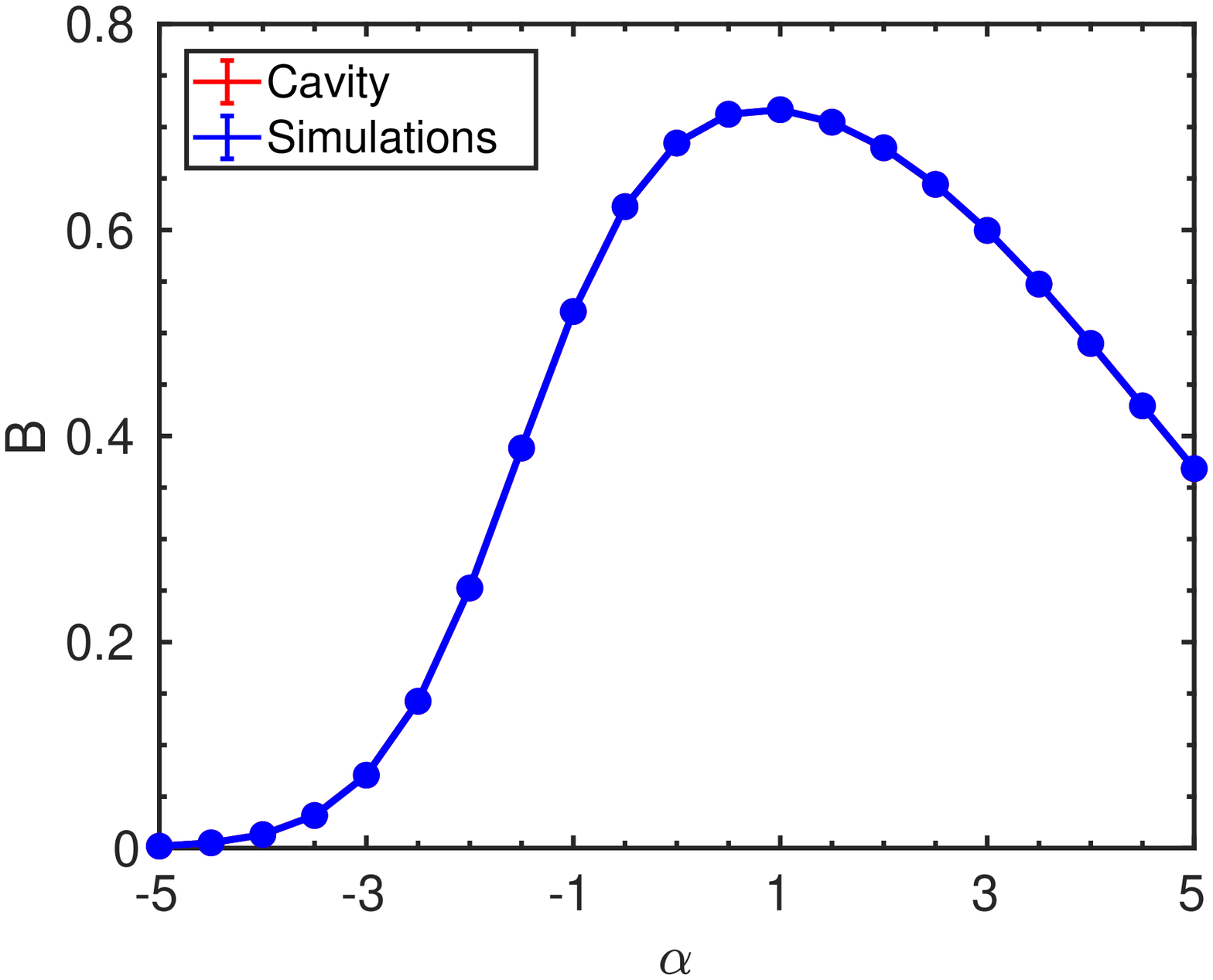}
\hfil
\includegraphics[width=0.475\textwidth]{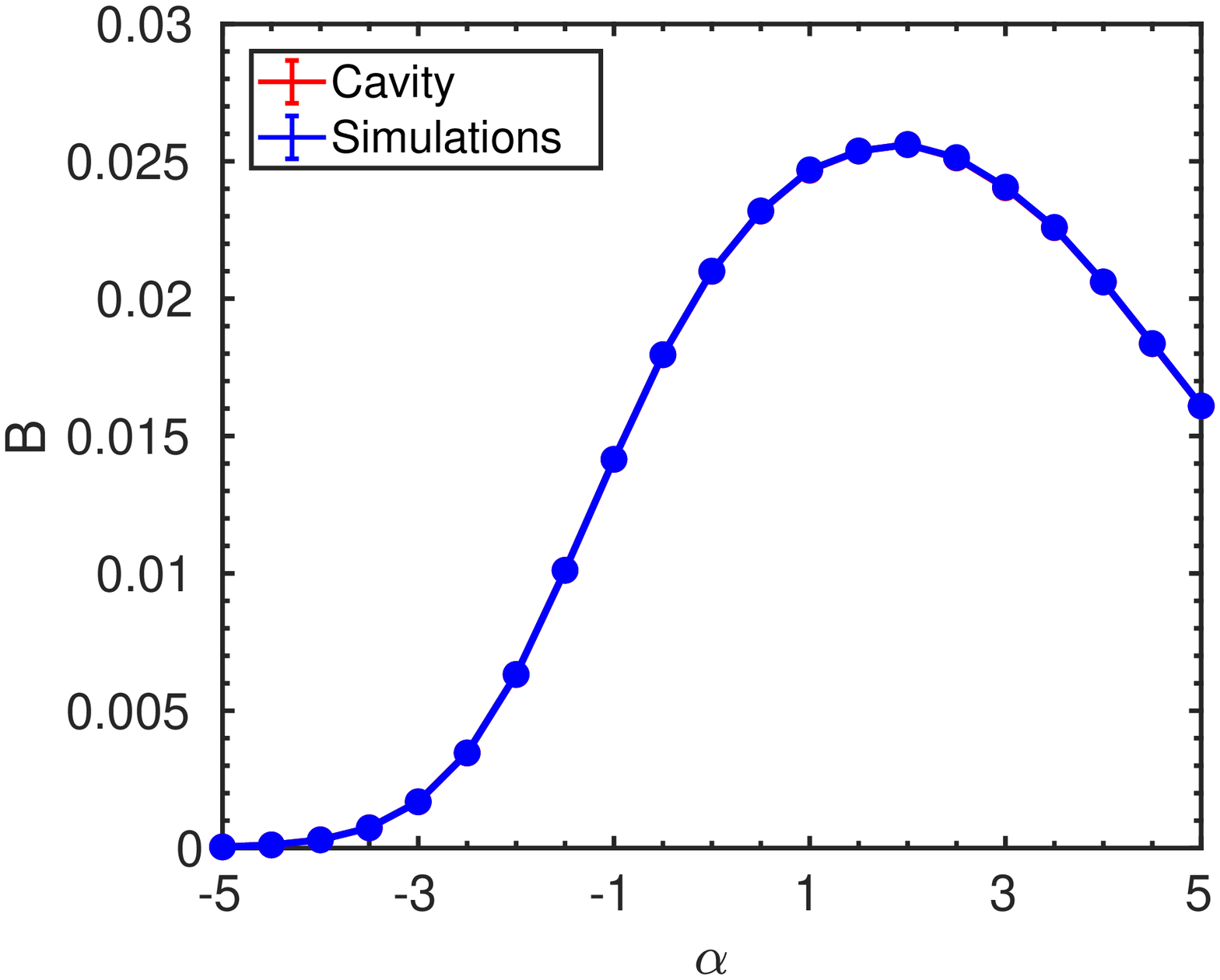}
\caption{Left panel: Network exploration efficiency $B$ of a power-law degree biased random walk $s(k)=k^\alpha$ 
on \er graphs of mean degree $c=4$. Right panel: Search efficiency of a power-law degree biased random walk computed 
for power-law degree biased hiding with $h(k) = k$ for the case where a fraction $\rho_h = 0.025$ of sites have an item 
hidden on them. The connecting line is a guide to the eye. On the scale of the figure, results obtained from the cavity 
method are indistinguishable from those obtained from random walk simulations.  The cavity results were obtained for 
giant components of systems of size $N=6000$, averaged over $N_s=2000$ random graphs.} 
\label{SvsCavity2}
\end{figure}

In Fig. \ref{SvsCavity2}, we compare the results of simulations with those obtained from
the cavity analysis for degree-biased random walkers with power-law degree bias $s(k)=k^\alpha$ 
for a range of $\alpha$ values between $\alpha=-5$ and $\alpha=+5$, and we observe very good 
agreement between the two. The cavity method can therefore be safely used as a substitute of 
random walk simulations for computing exploration and search efficiencies. In Fig. \ref{SvsCavity2} 
and below the symbols show the measured $B$ values, while the connecting lines are guides to the 
eye. Errors of both simulation and cavity results are estimated to be $\mc O(10^{-4})$ for the 
exploration efficiencies presented in Fig. \ref{SvsCavity2}, and $\mc O(10^{-5})$ for the search 
efficiencies, so error bars are mostly significantly smaller than the symbols indicating $B$ values. 
The same is true for results presented in the remainder of this paper.

The $\alpha$ dependence of $B$ can be understood by noting that very negative $\alpha$ will force 
the walker to spend most of her time at low degree sites, which are themselves surrounded by low 
degree sites, i.e., at the end of dangling chains in the graph, whereas very large positive 
$\alpha$ will entail that the walker is very likely to be found on sites with very high degrees 
that are themselves surrounded by high-degree nodes. Both extremes would prevent efficient 
exploration of the network, so large values of $B$ are expected at intermediate $\alpha$.

\begin{figure}[h]
\centering
\includegraphics[width=0.475\textwidth]{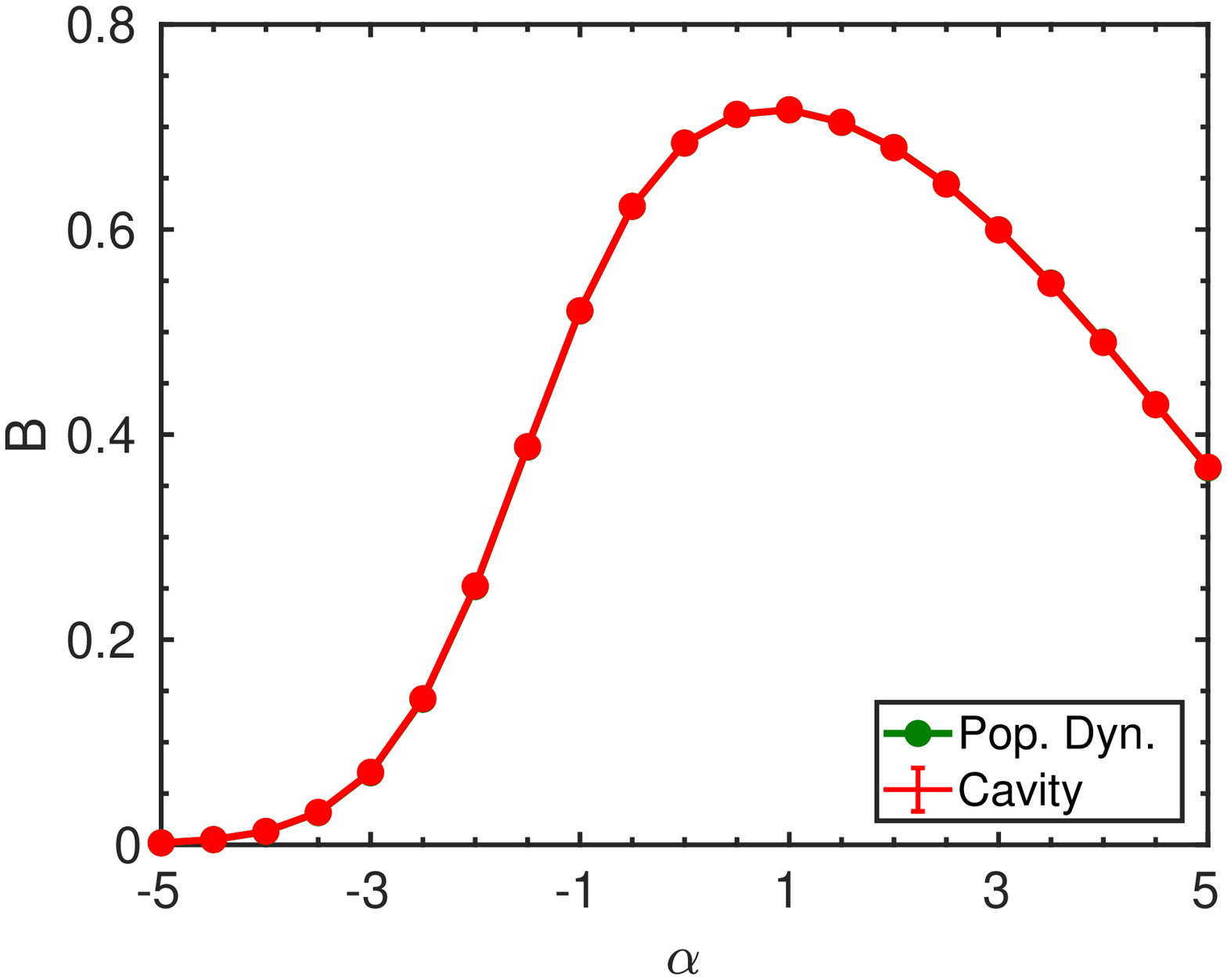}
\hfil
\includegraphics[width=0.475\textwidth]{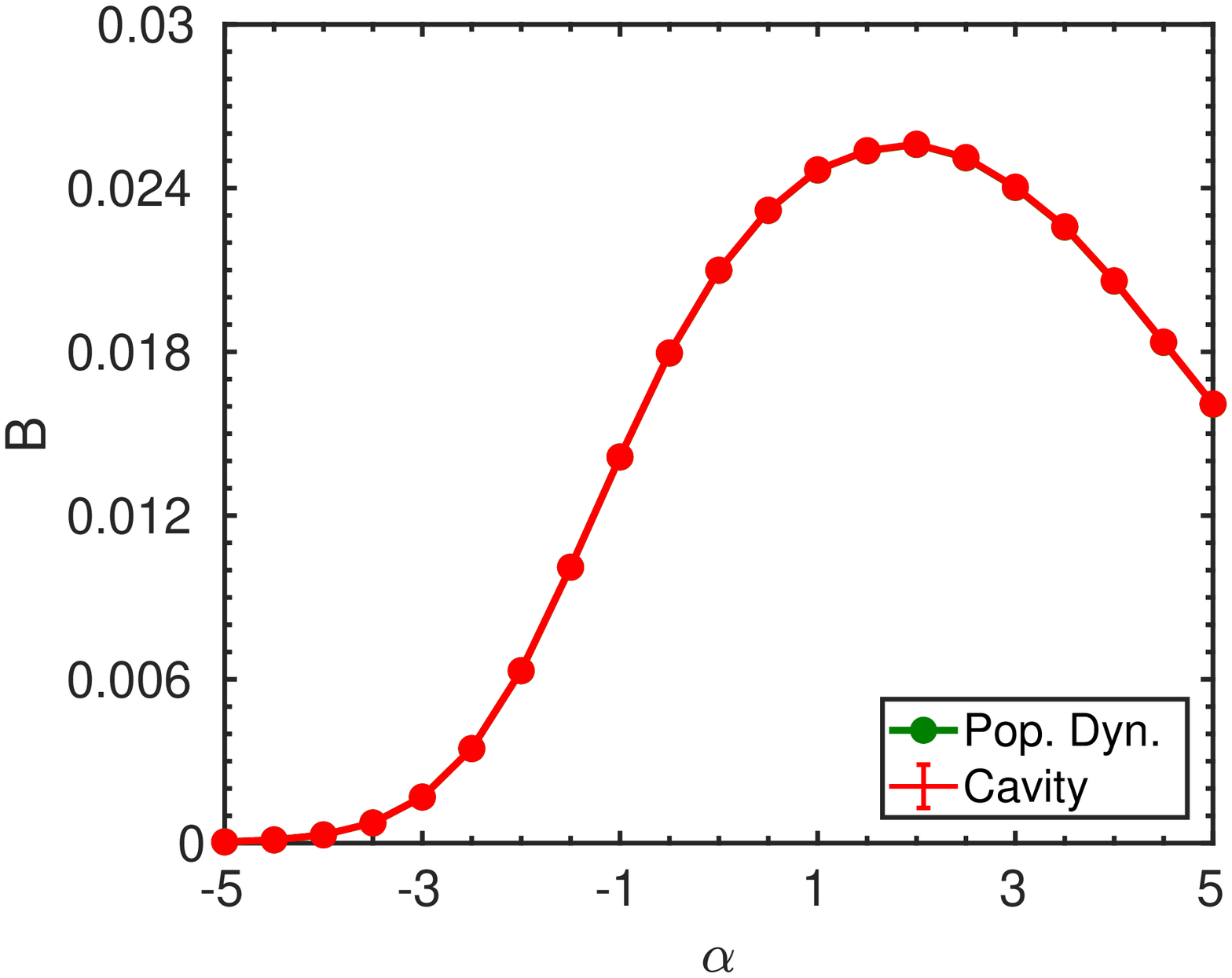}
\caption{Comparison of cavity and thermodynamic limit results for power-law biased random 
walk $s(k)=k^\alpha$ on \er networks of mean degree $c=4$. Left panel: Network exploration 
efficiency computed for $\xi_i\equiv 1$. Right panel: Search efficiency computed for degree 
biased power-law hiding with $h(k) = k$ for the case where a fraction $\rho_h = 0.025$ of 
sites has an item hidden on it. The cavity results were obtained for giant compontents of 
systems of size $N=6000$, averaged over $N_s=2000$ random graphs.} 
\label{PDvsCavity}
\end{figure}

In Fig. \ref{PDvsCavity}, finally we compare results of the single-instance cavity approach 
performed on the giant component of random graphs with those obtained using the theory for 
the thermodynamic limit. We see in Fig. \ref{PDvsCavity} that there is an excellent agreement 
between results obtained via averaging cavity results over single large problem instances and 
results obtained in the thermodynamic limit, provided the projection onto the giant component 
described in Sect. \ref{sec_gc} is performed. 

If one were to perform simulations by randomly selecting a starting vertex from the {\em entire 
system\/}, the starting vertex would belong to the giant component with probability $\rho$, whereas 
with probability $1-\rho$ it would belong to one of the finite clusters of the system. The contribution
of the latter to search and exploration efficiencies is zero, so one would expect average efficiencies 
for the entire system to obey
\be
B = \rho B_g + (1-\rho) B_f = \rho B_g\ ,
\ee
with $B_g$ and $B_f$ denoting search and exploration efficiencies corresponding to the giant component and
the finite clusters of the system, respectively.

Na\"ively applying the thermodynamic limit theory of Secs. 3.2.1 and 3.2.2 does {\em not\/} produce this
result (nor even $B\propto B_g$ with a proportionality constant that is independent of the search-strategy). 
The reason for this is that one of the key assumptions underlying the evaluation of search and exploration
efficiencies along the lines described in Sect. \ref{sec_theoryB}, viz. the fact that the Perron-Frobenius 
eigenvalue of the transition matrix is unique, ceases to be valid when the system contains several clusters
and the random walk transition matrix is thus decomposable.

\subsection{Hide and Seek}
\label{secHandS}
We now look at pitting different hiding and searching strategies against each other.
The main questions to be answered are concerned with identifying best search strategies
(within a given family), when pitted against hiding strategies (again within a given
family). Conversely, one might wish to identify the most efficient hiding strategy,
when pitted against given search strategies.

\begin{figure}[h!]
\centering
\includegraphics[width=8.cm]{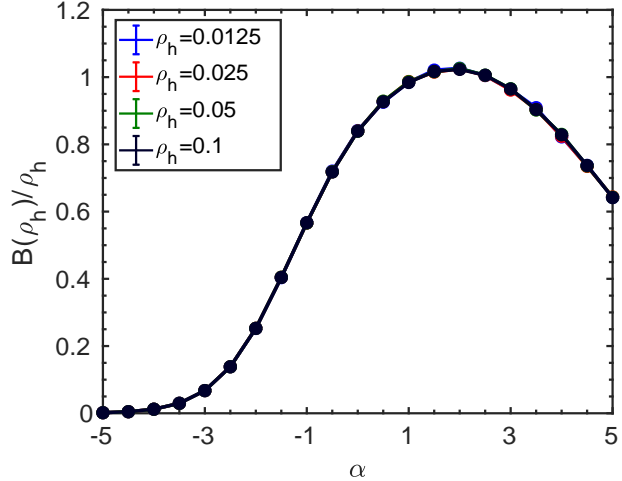}
\caption{Efficiency of power-law search with $s(k)=k^\alpha$ when set against power-law hiding 
of the form $h(k)=k^\beta$ with $\beta=1$. Shown are the ratios $B(\rho_h)/\rho_h$ for various 
values of $\rho_h$ in the allowed range defined by Eq. (\ref{rhoh-limit}), obtained by the single 
instance cavity method for the giant component of \er graphs with $c=4$ and $N$=6000, averaged over 
$N_s=2000$ instances. Curves lie on top of each other, verifying the expected proportionality.
}
\label{compBrhoh}
\end{figure}

Before presenting those results, let us point out though that the probability of hiding
items in any of the degree biased hiding strategies is according to Eq. (\ref{Exi1ofk})
proportional to the overall fraction $\rho_h$ of sites with an item hidden on them. It
is therefore expected, and explicitly borne out by Eqs. (\ref{Blim}) and (\ref{Blimg})
that search efficiencies in the large system limit will be proportional to $\rho_h$. We
verify this explicitly in Fig. \ref{compBrhoh} by displaying the ratio $B(\rho_h)/\rho_h$
as a function of search parameter $\alpha$ for power-law search pitted against a degree
biased hiding strategy of the form $h(k)=k$. Note that $B(\rho_h)/\rho_h > 1$ for the 
optimal $\alpha$ value, implying that the searcher is able to exploit the degree bias
of the hider to locate hidden items more effectively than expected by the fraction of
sites with items hidden on them. Unless stated otherwise we have in what follows chosen
$\rho_h =0.025$ for the density of hidden items.

\begin{figure}[t!]
\centering
\includegraphics[width=0.475\textwidth]{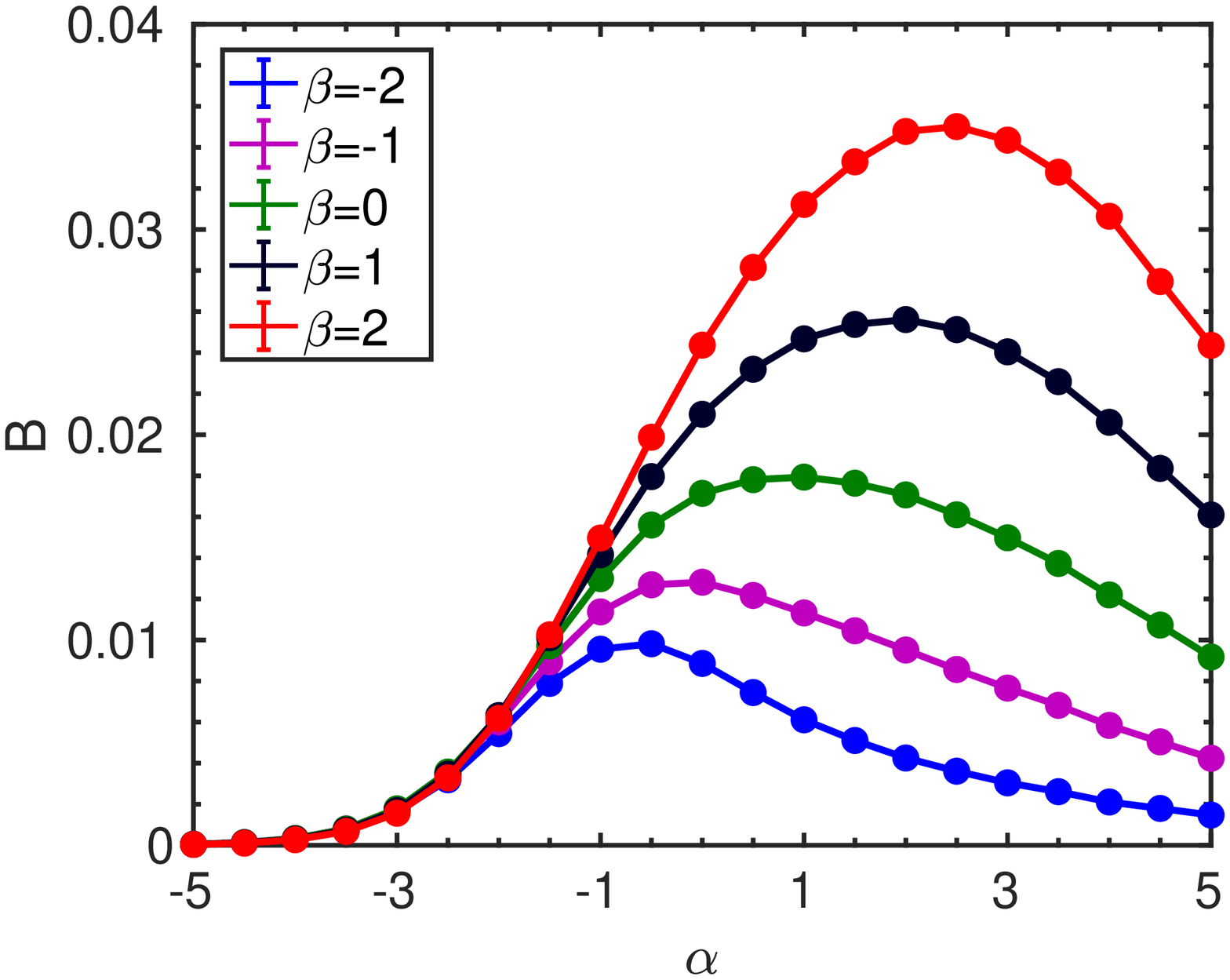}
\hfill
\includegraphics[width=0.475\textwidth]{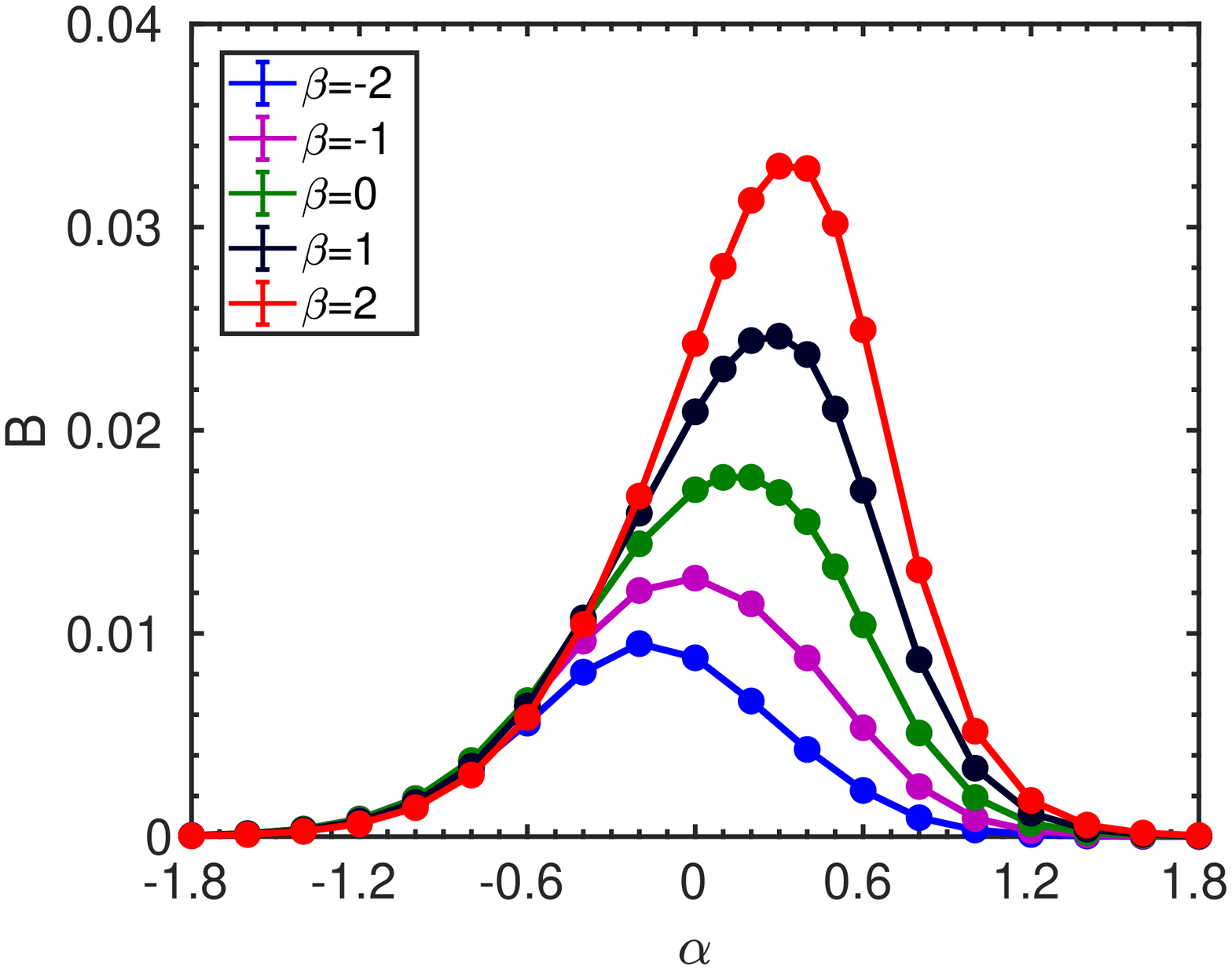}
\caption{Efficiency of power-law search with $s(k)=k^\alpha$ (left panel) and of exponential 
search with $s(k) = {\rm e}^{\alpha k}$ (right panel) as functions of $\alpha$, when set 
against power-law hiding of the form $h(k)=k^\beta$ for various $\beta$, and $\rho_h =0.025$. 
In both panels, curves from bottom to top correspond to increasing values of the bias parameter
$\beta$ of the hiding strategy. Shown are single instance cavity results for the giant component 
of \er graphs with $c=4$ and $N$=6000, averaged over $N_s=2000$ instances.}
\label{SKvsPL1}
\end{figure}

In Figure \ref{SKvsPL1} we investigate the search efficiency of power-law search (left
panel) and of exponential search (right panel), when pitted against power-law hiding.
We observe that there are optimal values of parameters of the search strategy which depend
on the exponent characterising the power-law hiding strategy. Optimal search efficiencies
are comparable in both cases, though matched functional forms for the degree bias of hiding
and searching generally perform slightly better than unmatched forms. The range of reasonably
effective search parameters is narrower for the exponential family. This is easily understood
as, for a given value of the bias parameter, exponential bias is generally more efficient in 
creating heterogeneity of transition rates than power-law bias.

\begin{figure}[t]
\centering
\includegraphics[width=0.475\textwidth]{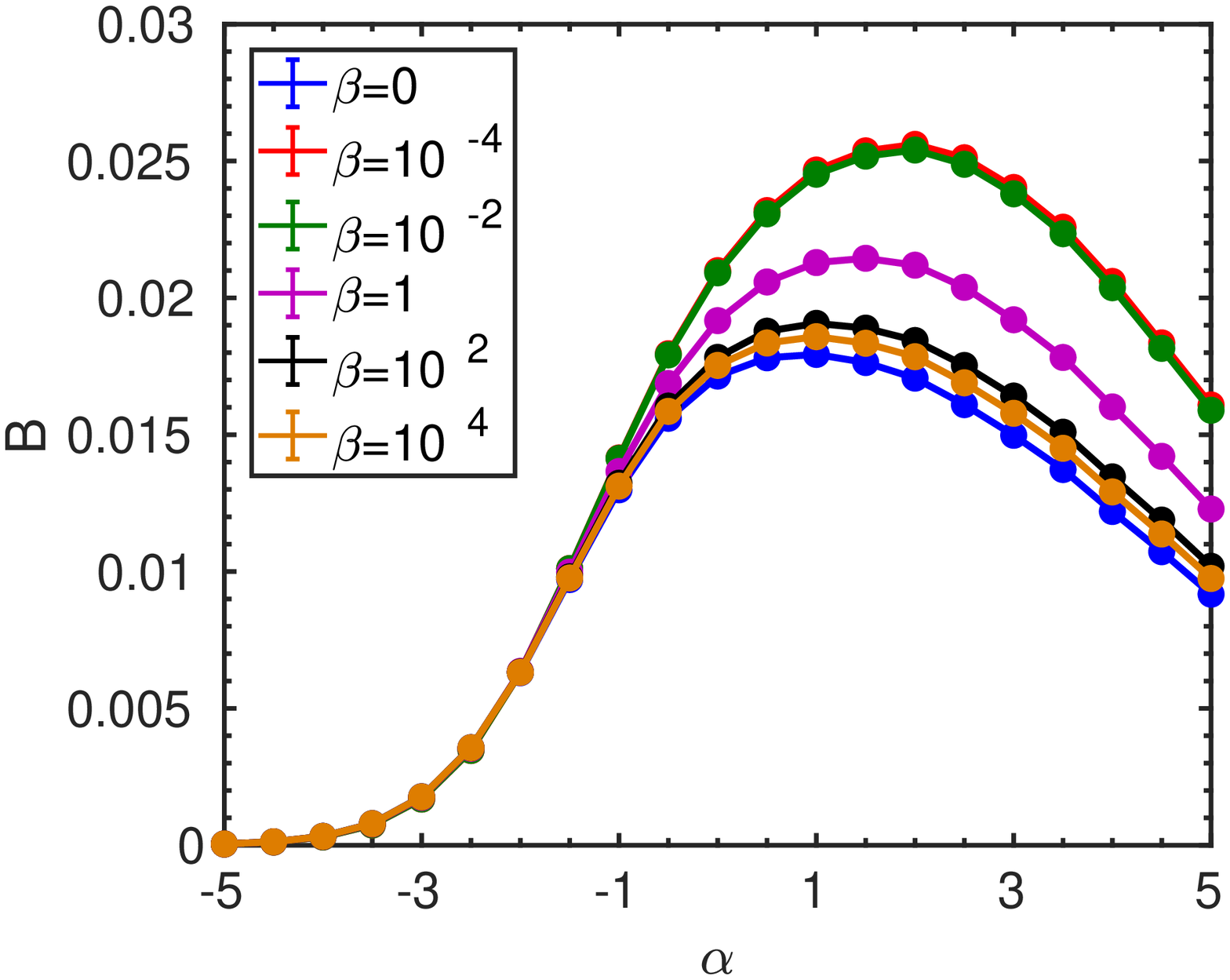}
\hfill
\includegraphics[width=0.475\textwidth]{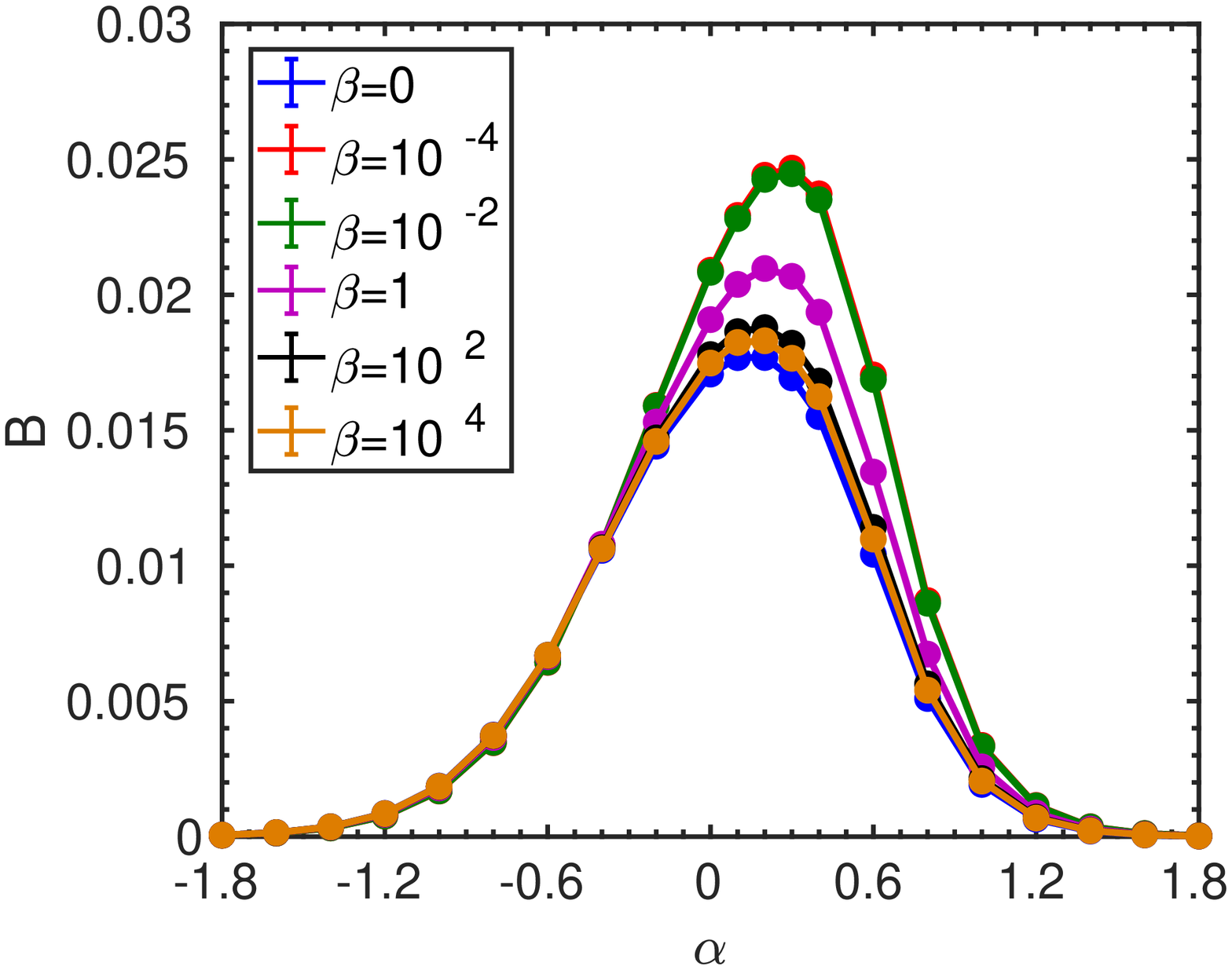}
\caption{Efficiency of power-law search with $s(k)=k^\alpha$ (left panel) and  exponential search 
with $s(k) = {\rm e}^{\alpha k}$ (right panel) set against logarithmic hiding of the form $h(k) = 
\log(1+\beta k)$ for various $\beta$, and $\rho_h =0.025$, with $\beta=0$ meant to refer to unbiased 
random hiding. In both panels, curves from bottom to top correspond to increasing values of the bias 
parameter $\beta$ of the hiding strategy. Shown are single instance cavity results for the giant 
component of \er graphs with $c=4$ and $N$=6000, averaged over $N_s=2000$ instances.}
\label{SKvsLog1}
\end{figure}

Fig. \ref{SKvsLog1} displays the efficiencies of power-law search (left panel) and exponential 
search (right panel), when set against logarithmic hiding of the form $h(k) = \log(1 + \beta k^{\gamma_h})$ 
with $\gamma_h=1$. In this figure we use the convention that $\beta=0$ is meant to refer to unbiased hiding. 
Note that in both cases the searcher's efficiency is always larger for degree-biased logarithmic hiding 
than for the unbiased hiding strategy with $\bt=0$.

\begin{figure}[h!]
\includegraphics[width=8.0cm]{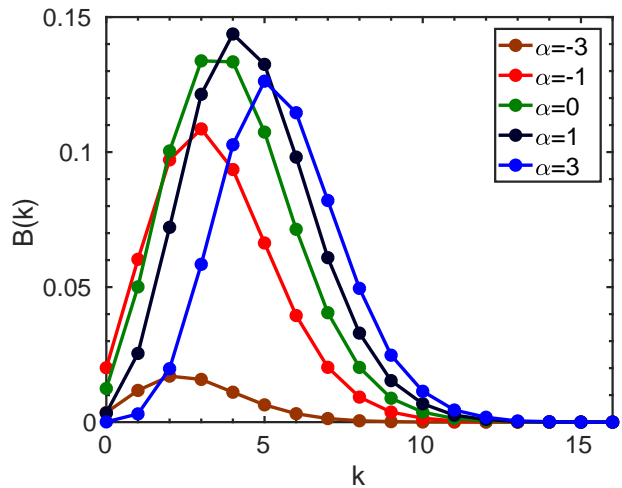}
\centering
\caption{Decomposition of the network exploration efficiency $B$ into $k$-dependent contributions 
for a range of bias parameters of the random walker with power-law bias of the form $s(k)=k^\alpha$. 
Shown are result obtained using population dynamics for the giant component of \er graphs with 
mean degree $c=4$.}
\label{decomposition}
\end{figure}

Fig. \ref{decomposition} illustrates the decomposition of exploration efficiencies according to  
Eq. (\ref{Bpk_search_n}) into contributions $B_k$ of vertices of different degree $k$ encountered 
in a degree biased walk with power-law degree bias of the form $s(k)= k^\alpha$. Peak positions 
indicating the degrees of sites which give the largest contributions to network exploration efficiencies 
are increasing functions of the bias parameter $\alpha$ of the random walker. Peak heights vary with 
$\alpha$, with the largest peak height corresponding to the optimal exploration bias $\alpha\simeq 1$
as observed in Fig. \ref{SvsCavity2}.

\begin{figure}[t!]
\includegraphics[width=8.0cm]{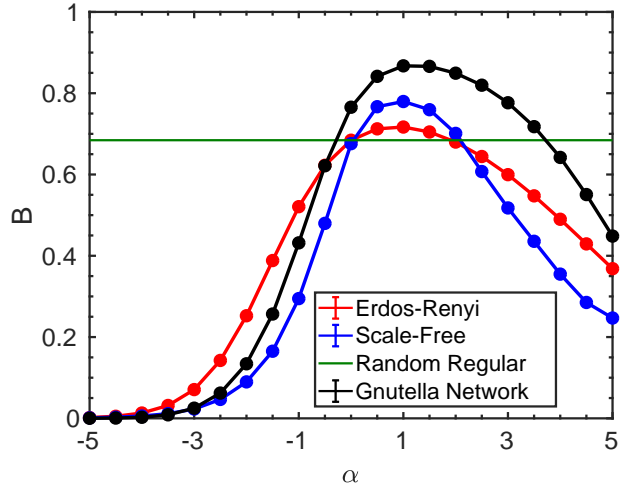}
\centering
\caption{Comparison of network exploration efficiencies for four different graph types 
using the cavity method. Parameters are $N$=6000, and $c$=4 for \er and regular random 
graphs; for the scale-free graph we chose $\gm=2.65$, with $k_{\rm min} =2$, $k_{\rm max} =400$ 
giving a mean connectivity $c=3.905$. Finally the Gnutella Network is a peer-to-peer file sharing 
network \cite{snap}, consisting of $N=36,682$ nodes, from which we have created an undirected 
version by symmetrising the links. Its aveage degree is $c=4.819$. The degree distribution of 
the Gnutella network exhihibits two regimes with distict power law behaviours, viz. $1 \le k 
\le 9$ where $p_k \propto k^{-1.74}$, and $11 \le k \le 40$ where $p_k \propto k^{-4.91}$. 
Curves with peak heights from bottom to top correspond to the Erd\H{o}s-R\'enyi, the scale-free 
and the Gnutella network, respectively. As expected there is no effect of degree bias on the 
exploration efficiency for the random regular graph.} 
\label{GraphTypes}
\end{figure}

\begin{figure}[b!]
\centering
\includegraphics[width=0.475\textwidth]{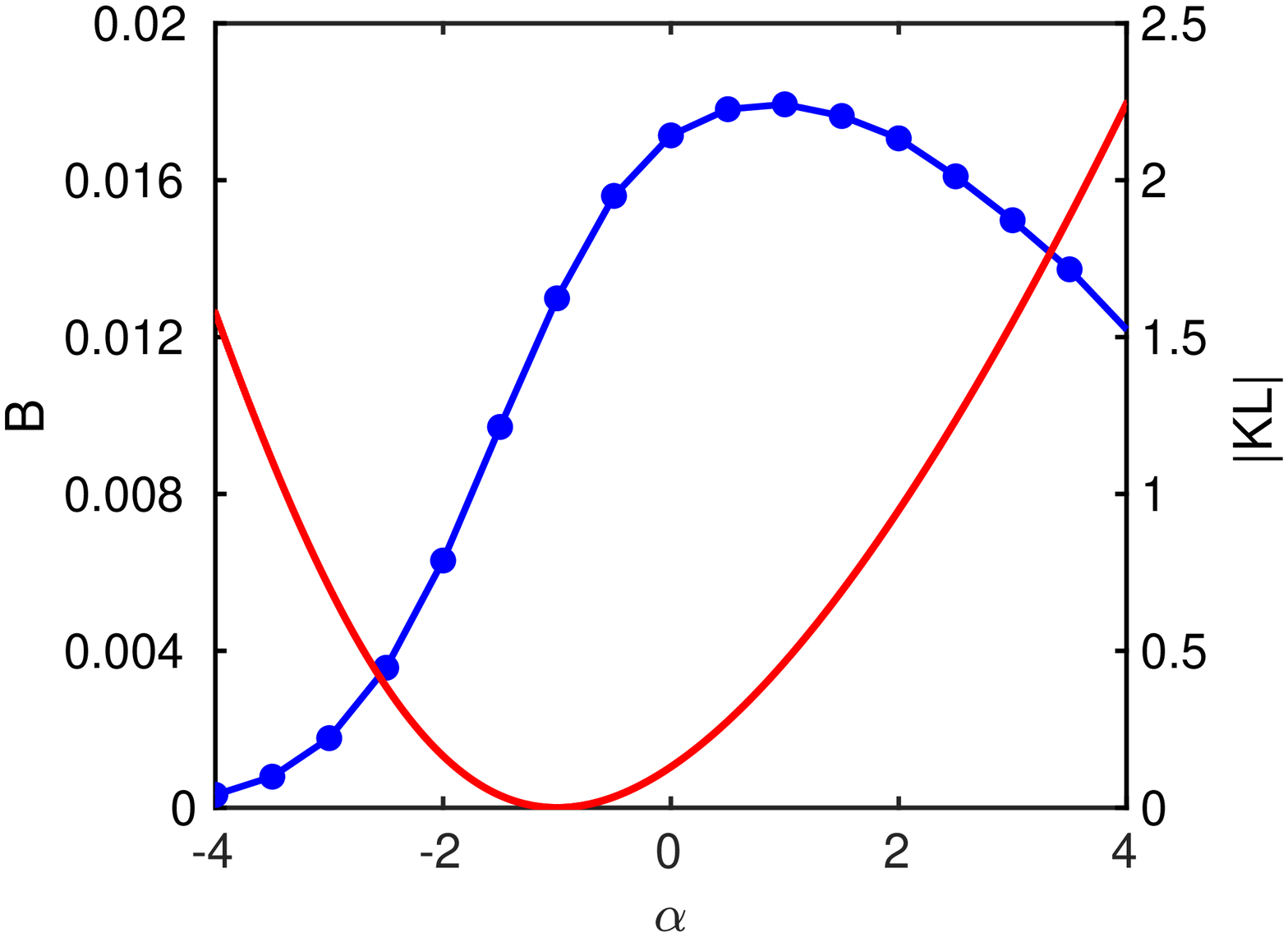}
\hfil
\includegraphics[width=0.475 \textwidth]{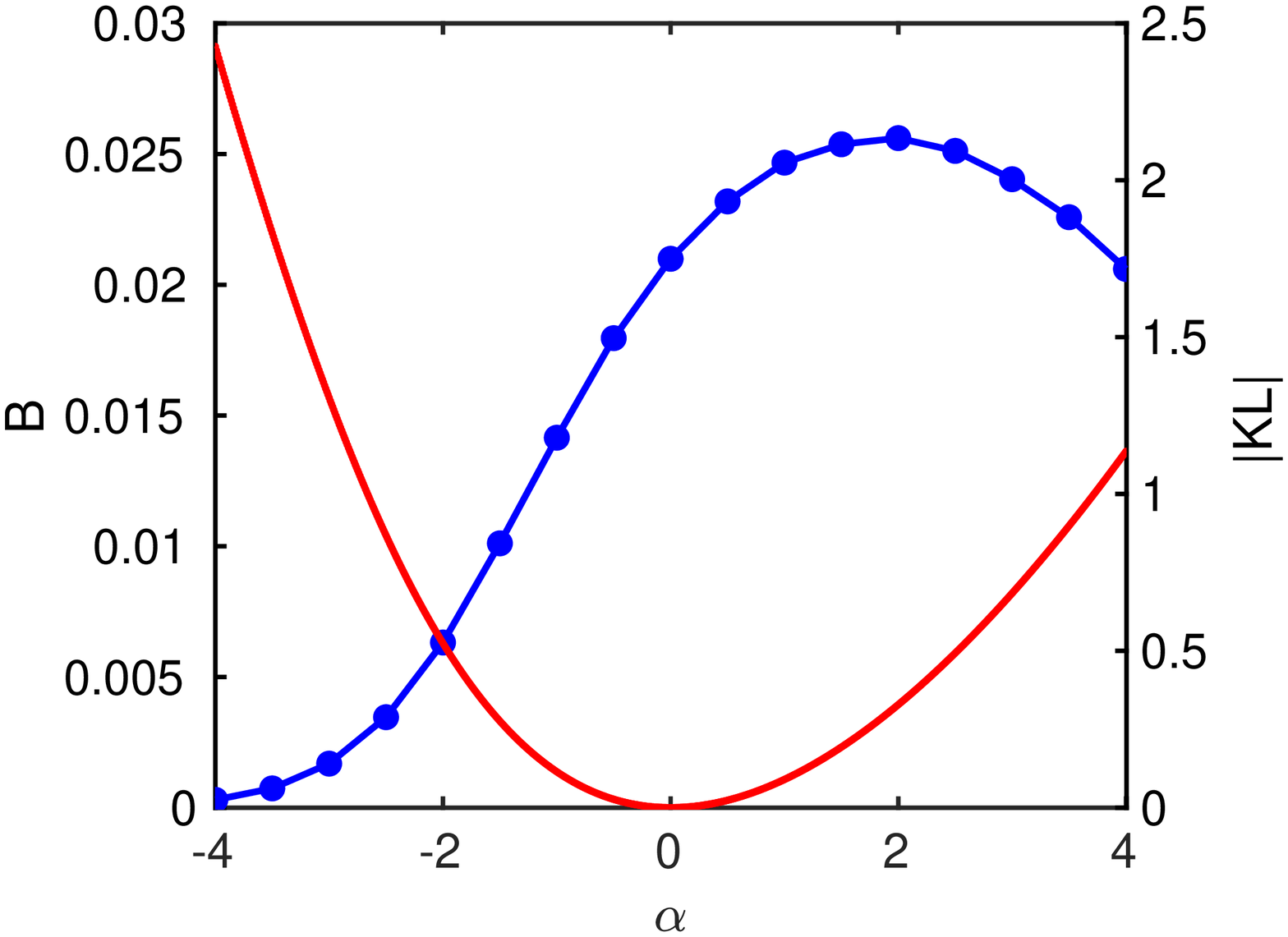}
\caption{Search efficiency and KL divergence displayed as functions of the bias parameter
$\alpha$ for power-law search $s(k)=k^\alpha$, set against random hiding (left panel)
and power-law hiding $h(k) = k^\beta$, with $\bt=1$ (right panel). In both panels, values
of search efficiencies  are displayed on the left axis, and those for KL divergences
on the right axis. Search efficiencies were obtained using cavity for \er graphs of size 
$N$=6000, with $c$=4 and $\rho_h = 0.025$, averaged over $N_s=2000$ samples.}
\label{KLD}
\end{figure}

Finally we address the question of the influence of the graph type on search or exploration
efficiencies. In Fig. \ref{GraphTypes} we look at exploration efficiencies of a degree biased
random walker on Erd\H{o}s-R\'{e}nyi, scale-free, and random regular graphs, as well as on a
real-world network --- a symmetrized version of the Gnutella peer-to-peer file sharing network 
\cite{snap}. For the random regular graph, any form of degree bias is clearly ineffective 
and the exploration efficiency must obviously be independent of the value of a formal bias 
parameter, as indeed confirmed by the results. Results also confirm the analytic prediction 
$B=2/3$ obtained in Eq. (\ref{rrB2}) for the $c=4$ system. The Gnutella network has an average 
degree of $c=4.819$, slightly higher than the mean degree of the synthetic networks, yet close 
enough to make for a meaningful comparison. For this real-world network, we have performed 
additional tests to verify that results of the cavity analysis agree with those of numerical
simulations. We found the agreement to be better than fractions of a per-cent. Results indicate 
that the degree bias is more effective in enhancing the exploration efficiency in the scale-free 
graph and the Gnutella network than in the \er graph, which is presumably due to the greater 
heterogeneity of the degree distributions in the scale-free system and the Gnutella network. 
For the latter though, the slightly higher mean connectivity may have further contributed to 
improved exploration efficiency at most $\alpha$ values.

\begin{figure}[t!]
\centering
\includegraphics[width=0.475\textwidth]{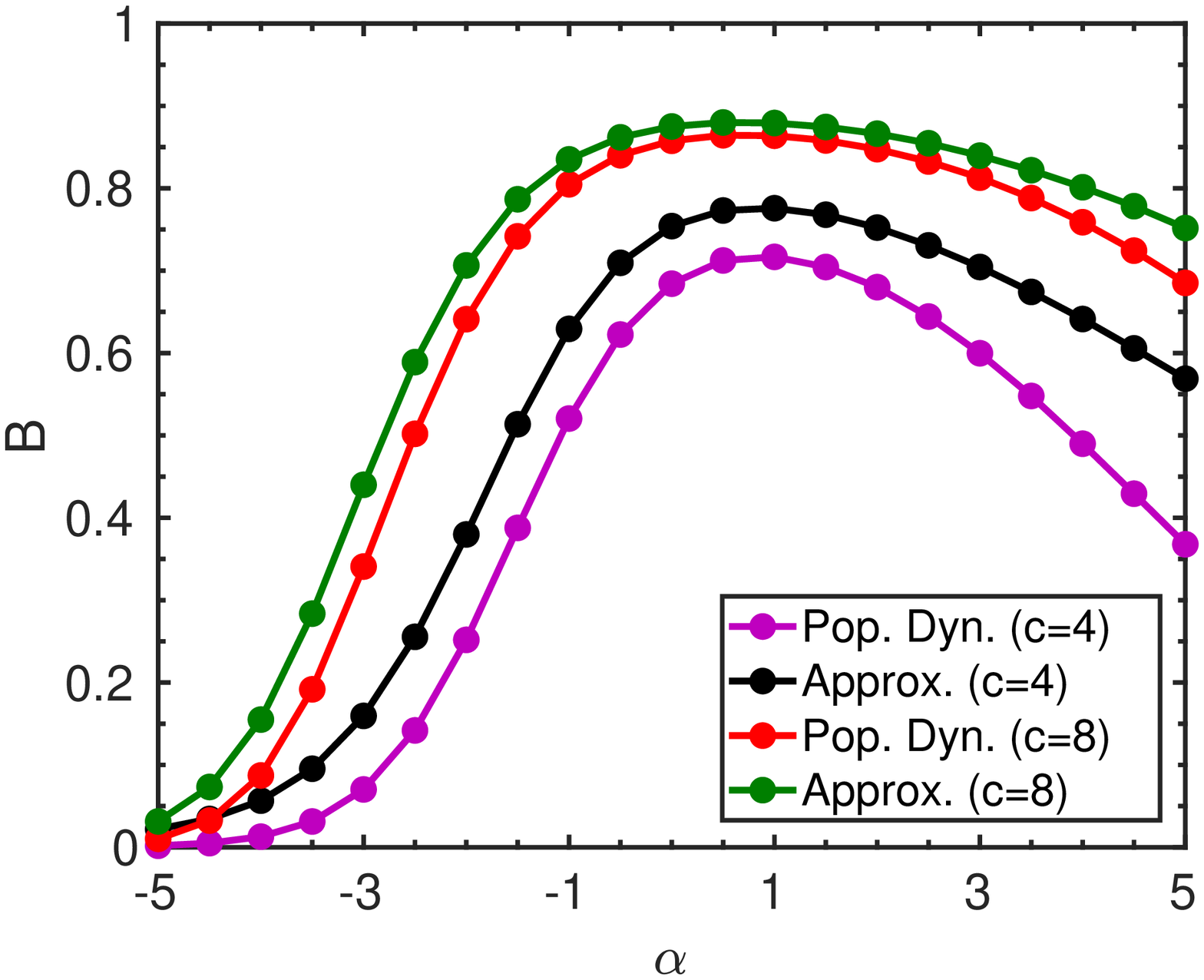}			
\hfill
\includegraphics[width=0.475 \textwidth]{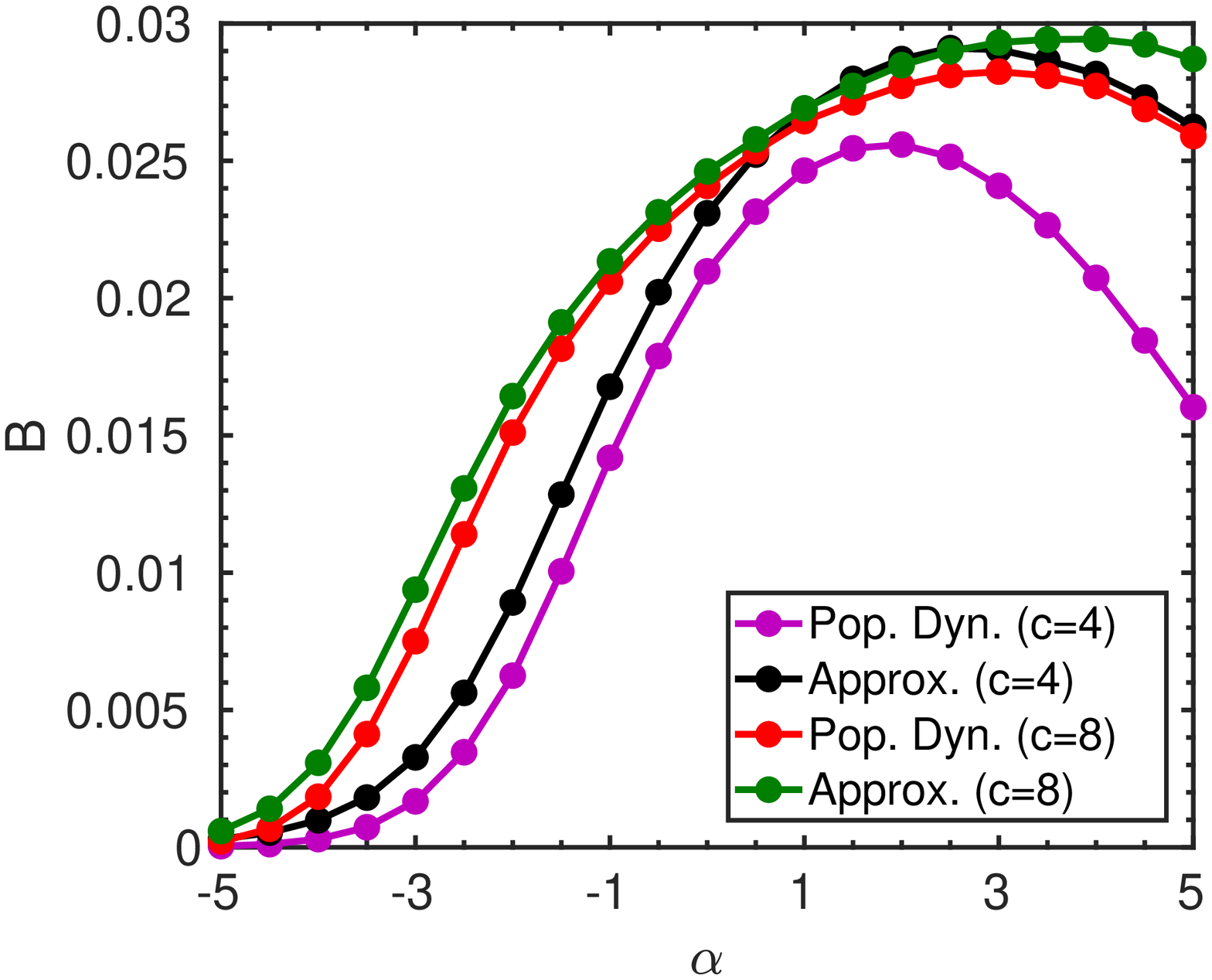}
\caption{Network exploration efficiency of a degree-biased random walker with degree bias 
following a power-law  $s(k)=k^\alpha$ as a function of the bias parameter $\alpha$ (left panel).
Efficiency of power-law search with $s(k)=k^\alpha$, set against power-law 
hiding $h(k) = k^\beta$, with $\bt=1$ as a function of the bias parameter $\alpha$ (right panel).
In both panels we compare results obtained via population dynamics for the thermodynamic limit 
with those of the non-backtracking approximation described in Sect. \ref{approx}, The upper pair
of curves in the left panel was computed for \er graphs of mean degree $c=8$, whereas the lower pair 
of curves shows results for \er graphs of mean degree $c=4$. The same trend is observed in the 
right panel, except for a small range of positive $\alpha$ values where the non-backtracking 
approximation predicts larger values of exploration efficiencies for the $c=4$ system than for 
the $c=8$ system.
}
\label{NonBT}
\end{figure}

We now turn to approximations. In Fig. \ref{KLD}, values of the KL divergence \eqref{Kl2} as
a function of the degree bias of the searcher are displayed together with the search efficiencies 
for the examples of power-law search set against random and power-law hiding.  While low values of 
the KL divergence are a reasonable qualitative predictor for high search efficiencies, the relation 
is not quantitative. In fact the minimum of the KL divergence occurs at a value of $\alpha$ which 
is approximately $\Delta \alpha \simeq 2$ {\em below\/} the value for which the search efficiency 
is maximised.

A discrepancy between the bias values which minimise the KL divergence and which maximise 
the values of the search efficiency is of course not unexpected, as the KL divergence
is based on equilibrium considerations whereas the search efficiency is a manifestly
non-equilibrium measure, as it doesn't account for (the frequency of) multiple visits 
of any given site, which is a characteristic equilibrium property.

In Fig. \ref{NonBT} we investigate the power of the non-backtracking approximation for network
exploration and search efficiencies. We expect this approximation to be efficient in networks in which
there are very few sites with low degrees, such as large mean degree \er graphs. Our results show
that the non-backtracking approximation is highly efficient qualitatively in that it predicts optimal
search and exploration parameters very accurately already for a $c=4$ \er graph. While the actually
predicted search and exploration efficiencies are for such a low mean degree system still 
$\mathcal{O}(25\%)$ off the mark, the approximation improves markedly across the entire $\alpha$ range
studied in the system with a (still moderate) mean connectivity $c=8$. The approximation is therefore
remarkably powerful, given that it is fairly straightforward, and in fact conceptually and technically 
{\em much\/} simpler than the full solution.

\section{Summary and Discussion} \label{conclusions}
We have studied the efficiency of random search strategies to locate items hidden
on a subset of vertices of complex networks, using a random walk framework. We 
assumed that items are hidden according to stochastic, degree biased strategies. 
In order to evaluate search efficiencies we adapt a result of De Bacco et al. 
\cite{debacco2015average} in which the average number of different vertices of 
a complex network visited by random walker performing an unbiased $n$-step random 
walk is computed, generalising their work by considering more general degree
biased transition probabilities. 

We use the cavity method to compute diagonal elements of resolvents needed for the 
evaluation of network exploration and search efficiencies for large single problem
instances. We also derive results for search efficiencies valid in the thermodynamic
limit $N\to\infty$ of infinite system size. This requires the solution of a degree 
dependent family of non-linear integral equations for inverse cavity variances. 
Their solution is obtained using a suitably adapted version of the population 
dynamics algorithm of M\`ezard and Parisi \cite{mezard2001bethe}.

It turns out that the na\"ive derivation, based on simply re-interpreting finite-instance 
self-consistency equations for inverse cavity variances as stochastic recursions 
in the thermodynamic limit does {\em not\/} accurately capture results valid for the 
giant component in the thermodynamic limit. The theory needs to be supplemented by 
degrees of freedom capturing whether sites do or do not belong to the giant cluster, 
as proposed in \cite{kuehn2016} in the context of sparse random matrix spectra. 

With this amendment, we find that results obtained using the cavity method for the 
giant components of large single instances of size $N=6000$ are already in excellent 
agreement with those obtained for the giant cluster in the thermodynamic limit, and 
both are in turn in excellent agreement with those of stochastic random walk simulations.

We found that search and network exploration efficiencies have a natural decomposition
in terms of degrees of sites contributing to the overall result, and we provided 
such a deconvolution of the network exploration efficiency by degree for the example 
of a random walker with a power-law degree bias. It is fairly easy to see that this
type of deconvolution could be carried out beyond degree, thereby identifying local
environments (such as degree and the collection of degrees in the first coordination
shell) that are most conducive to network exploration or search.

We have looked at various parameterised families of degree biased search algorithms and 
degree biased hiding strategies, namely power-law, exponential, and logarithmic families, 
as described in Table 1. Whatever the hiding strategy, we find --- for each family 
of search strategies --- a unique intermediate value of the parameter characterising
the search strategy that can be considered as optimal in the sense that it maximises
the search efficiency. An analogous statement can be made for the network exploration
efficiency. Qualitatively this could be understood by recalling that extreme values of 
the search parameter tended to imply that a degree biased random walker would spend most 
of her time at very low or very high degree sites, with both extremes not being 
conducive to efficient search or exploration.

We verified that search efficiencies were proportional to the density of items hidden
in the network, and we observed that normalised search efficiencies $B(\rho_h)/\rho_h$ 
could be larger than 1 if there was sufficiently strong degree bias in the hiding 
strategy which could be exploited to boost search.

Pitting matched and unmatched functional forms of hiding and search strategies against
each other, we always observed that the optimal search strategy in matched families 
slightly outperformed the most efficient search strategy when the functional forms of 
hiding and search strategies were unmatched.

We also used equilibrium dynamics consideration to locate efficient values for search
parameters by looking at the Kullback-Leibler distance between the distribution of
degrees the random walker visits in equilibrium and the distribution of degrees with
items hidden on them. The other approximation we looked into is a so-called non-backtracking 
approximation. It is based on the intuition that random walks on networks experience an
effective drift away from the starting vertex, which becomes very effective if vertices
typically exhibit large degrees. In such a situation one can  evaluate search efficiencies 
assuming that --- locally --- every non-backtracking step explores unseen parts of 
the network. We expect this approximation to be efficient in situations where networks
have few low degree sites, and we demonstrated that it was surprisingly effective 
on \er graphs even with modest mean degrees.

When comparing network exploration efficiencies for different graph types, we observed 
that, in an intermediate $\al$ range of power-law bias of the random walker, exploration 
on scale-free graphs is more efficient than on \er graphs, as a degree biased walker 
can make more efficient use of the large degree of heterogeneity of vertex degrees in 
scale-free graphs than of the rather small degree of heterogeneity present on \er networks. 
Large positive and large negative biases tend to localise the walker (near a set of nodes 
or sets of nodes) with either high or low degree vertices, leading to a reduction in 
exploration of search efficiencies. For the random regular graph, any degree bias is 
ineffective hence the search efficiency is found to be independent of the degree bias 
$\al$ as expected.

The hide and seek scenarios so far considered either did or did not have an item hidden on
a vertex. It is easy to see that efficiencies can be computed with graded values attached
to the items hidden on each site.

Our analysis has been restricted to computing {\em average\/} search and exploration
efficiencies. Clearly, from a security point of view discussed in the introduction it 
would be interesting to compute distributions of search efficiencies, in order to
assess, for instance, the likelihood of conducting unusually efficient, or unusually 
in-efficient searches. This would be particularly relevant if one were to insist 
locating all items hidden in the net and would thus have determine the cover-time 
\cite{cooper2005cover}, though not for the entire network but for a specified subset
of vertices. Such questions are for the time being outside the reach of our methods. 
They could, of course, always be addressed using simulations.

We have so far not dealt with proper game theoretical questions such as with the existence
and characterisation of Nash equilibria in the present problem, or with the possibility of
agents {\em learning\/} efficient search strategies, either on the fly or in repeated 
instances of the game. Analogous problems can be posed for the hider, who could update
their hiding strategy in repeated instances of the game, by observing the efficiency of
any strategy used by the seeker. 

We intend to address some of these questions in future publications.

\paragraph{Acknowledgement} The authors acknowledge funding by the UK Engineering and Physical 
Sciences Research Council (EPSRC) through the Centre for Doctoral Training “Cross Disciplinary 
Approaches to Non-Equilibrium Systems” (CANES, Grant Nr. EP/L015854/1). Insightful discussions
with Peter McBurney are also gratefully acknowledged.
 
\appendix
\section{Random Walk Analysis}
\label{app_RWF}
In what follows we briefly summarise the key elements of the derivation of  Eq. 
(\ref{hatrdiv}) which forms the basis of the evaluation of search efficiencies,
closely following \cite{debacco2015average}. 

The average number $S_i(\bsl {\xi},n)$ of marked sites visited in an $n$-step 
random walk, given by (\ref{eqSixi}), is expressed in terms of the probabilities 
$H_{ij}(n)$ of visiting node $j$ at least once in the first $n$ time steps when 
starting at node $i$. The $H_{ij}(n)$ can in turn be decomposed according to the 
time $m$ of the last visit to $j$ as
\be
H_{ij}\lp n \rp = \sum\limits_{m=0}^{n} G_{ij}(m)q_{jj}(n-m)\ ,
\ee
in which $ q_{jj}(n-m) $  denotes the probability for a walker who started at 
node $j$ not to return to node $j$ in $ n-m$ steps, and $G_{ij}(m)= (W^m)_{ij}$ is 
the $m$-step transition probability from $i$ to $j$. The convolution structure of 
the above expression entails 
\be
\hat{H}_{ij}(z) = \hat{G}_{ij}(z) \hat{q}_{jj}(z) \label{eq7}
\ee
for its $z$-transform. The $q_{jj}(n)$ are in turn related with first passage 
probabilities $F_{jj}(n)$ via 
\begin{align*}
q_{jj}(n-1)-q_{jj}(n) = F_{jj}(n)\ ,
\end{align*}
from which, with $q_{jj}(0)=1$ and $F_{jj}(0)=0$, one obtains
\be
\hat{q}_{jj}(z) = \frac{1-\hat{F}_{jj}(z)}{1-z} \ .
\ee
From
\be
G_{ij}(n) = \delta_{ij} \delta_{n0} + \sum\limits_{m=0}^{n} F_{ij} (m) 
G_{jj}(n-m)
\ee
finally one gets
\be
\hat{G}_{jj}(z) = \frac{1}{1-\hat{F}_{jj}(z)} \ , 
\ee
and thus
\be
\hat{H}_{ij}(z)=\frac{1}{1-z} \frac{\hat{G}_{ij}(z)}{\hat{G}_{jj}(z)} \ ,
\label{eq5a}
\ee
from which the $z$-transform of the number of items found in an $n$-step walk 
is obtained as
\be
\hat{S}_i(\bx, z)= \frac{1}{1-z} \sum\limits_{j \in \mc{V}} 
\frac{\hat{G}_{ij}(z)}{\hat{G}_{jj}(z)} \xi_j \ .
\ee
This is Eq. (\ref{eqhSixi}) in Sect. \ref{RWM}.

\section{Spectral Analysis}
\label{app_SPA}
To evaluate $\hat{S}_i(\bx, z)$ further one uses Eq. (\ref{GfromR}) and the spectral decomposition 
(\ref{Rhat}),
\be
\hat{R}(z) =  \frac{ \vv_1  \vv_1^T}{1-z} + 
\sum_{\nu=2}^N \frac{\vv_\nu \vv_\nu^T}{1-z \lambda_\nu} \equiv 
\frac{ \vv_1  \vv_1^T}{1-z} + \hat C(z) \ ,
\label{RhatApp}
\ee
of $\hat R(z)$. Using Eq. (\ref{firstevec}) for the components of the 
Perron Frobenius eigenvector and $\hat G_{jj}(z)= \hat R_{jj}(z)$ one has
\be
\hat{S}_i(\bx, z)= \frac{1}{1-z} \sum_{j \in \mc{N}} \Bigg[ 
\frac{s(k_j) \Gamma_j}{\hat R_{jj}(z) Y (1-z)} 
+ \frac{\sqrt{\frac{s(k_j) \Gamma_j}{s(k_i) \Gamma_i}}\hat C_{ij}(z)}
{\frac{s(k_j) \Gamma_j}{Y (1-z)} + \hat C_{jj}(z)}
\Bigg]\, \xi_j \ ,
\label{hatSbeforeLimits}
\ee
where, following \cite{debacco2015average}, we have used the spectral decomposition
(\ref{RhatApp}) of $\hat R_{jj}(z)$ in the denominator of the second contribution within
the square brackets in (\ref{hatSbeforeLimits}). Noting that $Y \propto N$, the second 
contribution can be argued to be negligible in the limit of large system size $N\to \infty$ 
and $z\to 1$ (in this order; see \cite{debacco2015average}, whereas the first contribution 
gives
\begin{eqnarray}
\hat{S}_i(\bx, z) &  \sim &  \frac{1}{(1-z)^2 Y} \sum\limits_{j \in \mc{V}} 
\frac{s(k_j)\gmm_j}{\hat{R}_{jj}}\,\xi_j\ ,\qquad z \to 1\ ,
\end{eqnarray} 
i.e. Eq. (\ref{hatrdiv}), where it is assumed that
\be
\hat{R}_{jj}  = \lim_{z \to 1}  \lim_{N \to \infty} \hat{R}_{jj}(z) = \lim_{z\to 1} \hat{C}_{jj}(z) 
\label{rhat1} 
\ee 
exists.

\section{Normalization Factors}
\label{app_Norm}
Here we describe the evaluation of the normalisation constants which appear in
the expressions for the search and exploration efficiencies $B$ in Eqs. (\ref{eq262n})
and (\ref{eq263}).

Looking at the normalisation constant $Y/N$ in Eq. (\ref{eq262n}), 
\be
\frac{Y}{N} = \frac{1}{N} \sum_{i \in \mc{V}}  s(k_i) \gmm_i\ ,
\ee
we evaluated it in the thermodynamic limit $N\to \infty$ as a sum of averages
by appeal to the law of large numbers (LLN). This gives $Y/N \to \mathcal{N}$ as $N\to\infty$,
with
\be
\mathcal{N} = \sum_k p_k s(k)\,\mathbb{E}\Big[\gmm_i \Big | k_i=k\Big]\ ,
\ee
 which is further evaluated as
\bea
\mathcal{N} &=& \sum_{k} p_k s(k) \, \mathbb{E} \Bigg[ \sum_{j \in \pr_i} \mathbb{E} \Big[ s(k_j) \Big | 
k_i =k \Big] \Bigg]\nn\\
  &=& \sum_{k} p_k s(k) \lb k  \sum_{k^\prm} p(k'|k) s(k^\prm) \rb =  
  \sum_{k} p_k s(k) \lb k  \sum_{k^\prm} \frac{\kp}{c} p_\kp s(k^\prm) \rb \ ,
\label{int1}
\eea
where we have used the fact that, for configuration model networks, the probability $p(\kp | k)$ 
that a site with degree $k'$ is adjacent to a site with degree $k$ does not depend on $k$, and 
is given by  $p(\kp | k) = \frac{\kp}{c} p_\kp$ in the last step. This is Eq. (\ref{cN}).

The evaluation of $Y_g/N_g$ in Eq. (\ref{eq263}) follows the same pattern, except for two 
crucial modifications. First, the degree distribution $p_k$ used above needs to be replaced by the 
degree distribution $p(k|n=1)$ conditioned on the giant cluster. Second, the giant cluster of a 
configuration model network is {\em not\/} a configuration model itself, so the conditional probability
$p(k'|k,n=1)$ that a vertex of degree $k'$ is adjacent to a degree $k$ site on a giant cluster does 
depend on the degree $k$. We can use results of Tishby et al. \cite{Tishby+18} who recently provided 
a comprehensive analysis of the micro-structure of the giant component of configuration model networks, 
including the two ingredients needed here.

We have
\be
  \frac{Y_g}{N_g}   = \frac{1}{N_g} \sum\limits_{i \in \mc{V}_g} {s(k_i)\gmm_i} \ ,
\ee
where $N_g$ is the size of the giant component. In the thermodynamic limit $N_g=\rho N\to \infty$
we have $Y_g/N_g \to \mathcal{N}_g$ by the LLN, where
\bea
\mathcal{N}_g &=& \sum_{k} p(k | n=1) s(k)\,\mathbb{E} \Bigg[\sum_{j \in \pr_i} \mathbb{E} 
\Big[s(k_j) \Big | k_i =k,n=1\Big]\Bigg]\nn\\
&=& \sum_{k} p(k | n=1) s(k)\, \Bigg[ k  \sum_{k^\prm} p(k'|k, n=1) s(k^\prm) \Bigg]\ .
\eea
Using
\be
p(k | n=1) =\frac{1}{\rho} \Big[1 -(1-\tilde \rho)^k\Big]\, p_k
\ee
and
\be
p(k'|k,n=1)  =  \lb \frac{1-{(1-\tilde{\rho} )}^{{k^\prm} -1}   
(1-\tilde{\rho} )^{k -1}  }{1 - (1-\tilde{\rho} )^k } \rb \frac{k^\prm}{c} 
p_k^\prm 
\ee
from \cite{Tishby+18}, in which $\rho$ is the percolating fraction, and $\tilde \rho$ is the probability
that a neighbour of a randomly selected vertex is on the giant component of the system, we obtain
\be
\mathcal{N}_g = \frac{c}{\rho}  \sum_{k,k^{\prime}} \frac{k }{c}p_k\   
\frac{k^{\prime} }{c} p_k^{\prime}\, s(k) s(k^\prm) \, \lb {1-{(1-\tilde{\rho} 
)}^{{k^\prm} +k -2}     } \rb \ .
\label{ap_pd2}
\ee
This is Eq. (\ref{gcnorm}).
\bibliography{HaSRW}
\end{document}